\documentclass[12pt]{article}

\usepackage{amsmath}
\usepackage{mathtools}
\usepackage[T1]{fontenc}
\usepackage{amssymb}
\usepackage{amsfonts}
\usepackage{amsthm}
\usepackage{mathrsfs}
\usepackage[all]{xy}
\usepackage{tensor}
\usepackage{stmaryrd}
\usepackage{bm}
\usepackage[labelfont={small, bf, sf}, textfont={small,sf}]{caption}
\usepackage{subcaption}
\usepackage{cite}

\usepackage[dvipsnames]{xcolor}
\definecolor{cadmiumred}{rgb}{0.89, 0.0, 0.13}
\definecolor{darkblue}{rgb}{0.2, 0, 0.8}
\definecolor{darkgreen}{rgb}{0.2, 0.71, 0}  
\definecolor{LinkGreen}{cmyk}{.9,0.05,.95,0.45}
\usepackage[colorlinks=true, 
citecolor=Blue,
linkcolor=Blue,
urlcolor=LinkGreen]{hyperref}

\numberwithin{equation}{section} 

\usepackage{ragged2e} 

\usepackage{authblk}
\usepackage{fullpage}

\newcommand{\beom}{\mathcal{E}}
\newcommand{\sub}{\mathcal{U}}

\newcommand{\lsq}{\left(l\cdot l\right)}

\newcommand{\p}[1]{  {\bm{ #1} }   }

\newcommand{\lie}{\pounds}
\newcommand{\ep}{\epsilon}
\newcommand{\vep}{\varepsilon}

\newcommand{\SSS}{\mathcal{S}}
\newcommand{\MMM}{\mathcal{M}}
\newcommand{\NNN}{\mathcal{N}}
\newcommand{\HHH}{\mathcal{H}}

\newcommand{\hateq}{\mathrel{\mathop {\widehat=} }}

\newcommand{\beq}{\begin{equation}}
\newcommand{\eeq}{\end{equation}}

\newcommand{\ind}[1]{\indices{#1}} 

\newcommand{\Hajicek}{H\'aj\'{i}\v{c}ek~}

\title{Anomalies in gravitational charge algebras of null boundaries
and black hole entropy}
\author{Venkatesa Chandrasekaran\thanks{ven\_chandrasekaran@berkeley.edu}}
\affil{\small \it 
Center for Theoretical Physics and Department of Physics,
University of California, Berkeley, CA 94720, U.S.A. }
\author{Antony J. Speranza\thanks{asperanz@gmail.com} }
\affil{\small \it 
Perimeter Institute for Theoretical Physics, 31 Caroline St. N, Waterloo, ON N2L 2Y5, Canada}
\date{}

\begin{document}

\maketitle

{\abstract
We revisit the covariant phase space formalism applied to gravitational theories with null boundaries, utilizing the most general boundary
conditions consistent with a fixed null normal.  
To fix the ambiguity inherent in the 
 Wald-Zoupas definition of 
quasilocal charges,
we propose a new principle, based on holographic reasoning, 
that the flux be of Dirichlet form.  This also produces an expression
for the analog of the Brown-York stress tensor on the null surface.  
Defining the algebra
of charges using the Barnich-Troessaert bracket for open subsystems, 
we give a general formula for the central---or more generally, 
abelian---extensions that appear in terms of the anomalous
transformation of the boundary term in the gravitational action.
This anomaly arises from having fixed a frame for the null normal, and we draw parallels between it and the holographic Weyl 
anomaly that occurs in AdS/CFT.  As an application of this formalism,
we analyze the near-horizon Virasoro symmetry considered 
by Haco, Hawking, Perry, and Strominger, and perform a systematic derivation 
of the fluxes and central charges.  
Applying the Cardy formula to the result yields
an entropy that is twice the Bekenstein-Hawking entropy of the horizon.
Motivated by the extended Hilbert space construction, we interpret this in terms of a pair of entangled CFTs 
associated with edge modes on either side of the bifurcation surface.
}

\flushbottom

\newpage

\tableofcontents

\section{Introduction and summary}

Observables in general relativity tend to be global in nature, 
owing to the fact that diffeomorphisms are gauge symmetries of the
theory.  This large gauge redundancy causes the Hamiltonian
of the theory to be localized to the asymptotic boundary, and 
diffeomorphism-invariant observables must be constructed 
relationally, using the fixed structures at the asymptotic boundary
as points of reference \cite{Giddings2006a, Marolf:2008mf, Donnelly_2016}.  
Nonetheless, there exist notions of quasilocal
observables that describe 
degrees of freedom inside of spatial subregions. In particular,
several approaches to understanding the origin of black hole entropy
deal with quasilocal charges on the event 
horizon \cite{Carlip_1999,  Bagchi2013, 
Hawking_2016, Donnelly2016a, Afshar2016, Carlip2018a, Haco2018, Aggarwal2020}. Moreover, charges associated with 
$\mathscr{I}$ in asymptotically flat space
\cite{Wald2000b, Barnich2010e, Barnich2010d, He2014b,
Strominger2018} and more general 
null surfaces \cite{CFP, Hopfmuller2017a, Hopfmuller2018, Grumiller2020a, Adami2020a, Adami2020b, Grumiller2020b} 
have received recent attention, due to their
potential relevance to quantum gravity and flat space holography.

The appearance of quasilocal observables when considering subregions
can be understood in terms of symmetry breaking.  The 
introduction of a fixed boundary partially violates the diffeomorphism
symmetry present in the theory, causing some transformations that 
were formerly considered gauge to become physical \cite{ Carlip_1999,Carlip1995}.  The charges associated 
with the broken diffeomorphisms localize on the boundary of the subregion, and hence are referred to as edge modes 
\cite{Donnelly2016a,Balachandran_1996, Speranza2018a}.  The connection to black
hole entropy comes from the proposal that the edge modes represent
the degrees of freedom counted by the Bekenstein-Hawking
entropy of a surface, given by $S_{\text{BH}} = \frac{A}{4G}$,
with $A$ the area of the surface.  The fact that the edge modes are localized 
on the boundary qualitatively explains the scaling with area, but 
in some examples the numerical coefficient can be computed in 
a precise manner.  As first shown by 
Strominger for BTZ black holes in $\text{AdS}_3$ \cite{Strominger_1998} using the Brown-Henneaux 
central charge \cite{Brown1986b}, 
and subsequently
generalized by Carlip to generic Killing horizons \cite{Carlip_1999,Carlip_1999a}, 
if the quasilocal charge algebra includes a Virasoro algebra,
the entropy can be derived by applying the Cardy formula
for the entropy of a 2D conformal field theory \cite{Cardy1986}.  
The rationale behind this procedure is that the Virasoro algebra
is the symmetry algebra of 2D CFTs, so it is natural to conjecture that
the quantization of the edge modes is given by a CFT, 
with the central charge determined by the classical brackets of the
quasilocal
charges.  The precise agreement between the Cardy entropy and 
the Bekenstein-Hawking entropy then provides {\it a posteriori} 
justification for associating the entropy with edge mode degrees of 
freedom. 

In most constructions in which the entropy arises from the 
Cardy formula applied to a boundary charge algebra, boundary
conditions are needed to ensure the charges are integrable.
The need for boundary conditions arises because the vector fields
generating the symmetry have a transverse component to the 
codimension-2 surface on which the charge is being evaluated.  
This means they are generating a transformation that moves the
bounding surface, and hence without boundary conditions, symplectic
flux can leak out of the subregion as the system evolves.  Imposing
the boundary conditions ensures that the subregion behaves as a closed
system, but gives the boundary the status of a physical barrier, 
preventing exchange of information between the subregion and its 
complement.  When viewing the boundary as an arbitrary partition
used to define a subregion, 
one would like a definition of quasilocal charges that does not
employ such restrictive boundary conditions, and need not require 
conservation under time
evolution.  In the place of conservation, 
one seeks an independent definition 
of the flux of the quasilocal charge through the subregion boundary, 
so 
that the charge instead obeys a continuity equation.  
For general relativity and other diffeomorphism-invariant theories,
Wald and Zoupas provided such a construction of quasilocal 
charges using covariant phase space techniques \cite{Wald2000b}, 
and its 
application to null boundaries at a finite location was 
considered in \cite{CFP}.

Another reason for utilizing the Wald-Zoupas prescription is that 
in some cases, there is no obvious boundary condition that ensures
integrability of the quasilocal charges.
Such was the situation encountered by Haco, Hawking, Perry, and Strominger (HHPS) \cite{Haco2018}, who 
identified a set of near-horizon Virasoro symmetries for Kerr black holes,
inspired by the  hidden conformal symmetry of the near horizon
wave equation identified in \cite{Castro2010}.
These symmetries suggest a possible extension of the results 
of the Kerr/CFT correspondence \cite{Guica2009, Compere2012},
which deals with extremal Kerr black holes, to a holographic
description of more general horizons.  
There does not exist a local boundary condition one can
impose on the dynamical fields that is preserved by the HHPS vector
fields, while simultaneously ensuring integrability of the 
corresponding charges.\footnote{There can be weaker, integrated 
boundary conditions that ensure integrability for special
choices of the parameters defining the transformation, as described
in \cite{Chen:2020nyh}.}  Hence,  the 
Wald-Zoupas procedure is needed to define the quasilocal charges. 

A specific form of the flux in the Wald-Zoupas prescription was 
conjectured in \cite{Haco2018}, and was also used in various subsequent
works generalizing the construction \cite{Aggarwal2020, Haco2019, 
Chen:2020nyh, Perry2020}.  
The goal of the present work
is to derive the necessary Wald-Zoupas prescription for these 
constructions from first principles.  
In order to do so, there are three main technical challenges that 
need to be resolved.  

First, there are a number of ambiguities that arise when carrying
out the Wald-Zoupas construction, some of which affect the final 
result for the entropy. The most important ambiguity is in the 
ability to shift the symplectic potential on the bounding
hypersurface by total variations, which subsequently affects the 
definitions of the charges and fluxes.  To resolve this issue,
we first reformulate the Wald-Zoupas procedure in section \ref{sec:quasicharge}
using Harlow and Wu's 
presentation of the covariant phase space formalism with boundaries \cite{Harlow2019}.  
Doing so allows for an efficient parameterization of the
ambiguities that can appear in terms of boundary and corner terms
in the variational principle.  Rather than imposing boundary conditions
to eliminate some terms that appear in the variations,
as was done in \cite{Harlow2019}, we interpret the nonzero 
boundary terms as representing a symplectic flux through the
boundary. Explicitly, we decompose the pullback $\boldsymbol{\theta}$ of the symplectic potential current into boundary $\ell$, corner $\beta$, and flux $\beom$ terms:
\begin{align}
    \boldsymbol{\theta} + \delta \ell = d\beta + \beom. \label{introdecomp}
\end{align}
Resolving the ambiguities in the Wald-Zoupas prescription
then amounts to finding a preferred choice for the flux term $\beom$.

We propose a principle for fixing this ambiguity
in section \ref{sec:quasicharge}, 
namely that $\beom$ should be of Dirichlet form,
meaning it involves variations only of intrinsic quantities
on the surface. It therefore is expressible as 
\beq
\beom = \pi^{ij}\delta g_{ij},
\eeq
where $\delta g_{ij}$ is the variation of the induced metric
on the bounding hypersurface, and $\pi^{ij}$ are the conjugate
momenta constructed from extrinsic quantities.  For null hypersurfaces, the variation of the null generator 
$\delta l^i$ is also considered an intrinsic quantity, so the Dirichlet
form of the flux  in this case reads
\beq
\beom = \pi^{ij}\delta g_{ij} + \pi_i \delta l^i.
\eeq
The terminology
``Dirichlet'' refers to the fact that vanishing flux is equivalent to Dirichlet boundary
conditions for this choice.  
The Dirichlet flux condition is a novel proposal in the context of 
the Wald-Zoupas construction, in contrast with previous proposals which 
employed properties of the flux in stationary solutions to partially 
fix its form \cite{Wald1993a, CFP}.  However, it is 
familiar from the Brown-York procedure for quasilocal energy
\cite{Brown:1992br}, and  has a natural interpretation in the context 
of holography.  We also argue that this form of the flux is preferred
from the perspective of gluing subregions together
in the gravitational path integral \cite{Carlip_1995}. As a byproduct of fixing this form of the flux, we 
can also employ Harlow and Wu's \cite{Harlow2019} resolution of the 
standard Jacobson-Kang-Myers ambiguities in the covariant phase 
space formalism \cite{Jacobson1994b, Iyer1994a}, 
leading to unambiguous definitions of 
the quasilocal charges.  

The second issue to address is the problem of constructing a bracket
for the quasilocal charges that defines their algebra. Poisson
brackets are not available when employing the Wald-Zoupas procedure,
since we are dealing with an open system with respect to the 
symplectic flux. Therefore, in section \ref{sec:BTbrack}, 
we instead utilize the bracket 
defined by Barnich and Troessaert in \cite{Barnich:2011mi}
for nonintegrable charges.  It has the advantage of representing 
the algebra satisfied by the vector fields generating the symmetry
transformations, up to abelian extensions.  We further show that 
the algebra extension has a simple expression 
\beq
K_{\xi,\zeta} = \int_{\partial\Sigma}
\Big(i_\xi \Delta_{\hat\zeta}\ell 
- i_\zeta \Delta_{\hat\xi} \ell\Big)
\eeq
 in terms of
$\Delta_{\hat\xi}\ell$,
the anomalous transformation with respect to the symmetry generator
$\xi^a$ 
of the boundary term $\ell$ in \eqref{introdecomp}. The anomaly operator  $\Delta_{\hat\xi}$, defined in (\ref{eqn:anomop}), 
directly measures the failure of an object to transform covariantly
under the diffeomorphism generated by $\xi^a$, and hence we 
immediately see that algebra extensions only appear when the 
boundary term $\ell$ is not covariant with respect to the
transformation.  Because the Barnich-Troessaert bracket coincides
with the Poisson bracket when the charges are integrable, this formula
for the extension applies in the case of integrable charges as well.  
This shows quite generally that central charges and abelian
extensions appear as a 
type of classical anomaly associated with the boundary
term in the variational principle.  This statement
is directly analogous to the appearance of holographic 
Weyl anomalies in AdS/CFT \cite{Henningson1998a, Balasubramanian1999a, DeHaro2001, Papadimitriou2005}.

The third issue to address is finding a decomposition of the 
symplectic potential for general relativity when restricted to 
a null boundary $\NNN$.  This question has been treated in previous
analyses \cite{CFP,Parattu_2016, Hopfmuller2017a, Hopfmuller2018, Oliveri:2019gvm, Lehner_2016}; 
however, most of these employ boundary conditions that are 
too strong to allow for the symmetries generated by 
the HHPS vector fields.  In our analysis in section \ref{sec:symppot}, we employ the weakest possible boundary conditions that 
ensure the presence of a null surface, and in which the variations
of all quantities are entirely determined in
terms of $\delta g_{ab}$. This is done by fixing the 
normal covector, $\delta l_a = 0$, and imposing nullness
by requiring that $ l^a l^b \delta g_{ab} = 0$ on $\NNN$.  
The covector $l_a$ is thus viewed as a background structure
introduced into the theory in order to define the boundary.  
Because it is a background structure, no issues arise if the 
symmetry generators do not preserve it; in fact, the failure 
of $l_a$ to be preserved by the symmetry generators is the sole
source of noncovariance in the construction, and hence is responsible
for the appearance of a nonzero central charge.  By contrast, it is 
crucial that the vector fields satisfy $l^a l^b \lie_\xi g_{ab}  = 0$ on
$\NNN$, since this arises from a boundary condition imposed on the 
dynamical metric; violating it would cause the symmetry transformations
to be ill-defined. The HHPS vector fields satisfy this condition,
as do any vectors which preserve the null surface.  

The result of the 
decomposition of the symplectic potential is given in
equations (\ref{decomp})--(\ref{eqn:pii}), 
in which the Dirichlet form of $\beom$ is decomposed into  
$\frac{d(d-1)}{2}$  
canonical pairs on the null surface.  The decomposition that 
we find has appeared before in \cite{Parattu_2016}, and 
related decompositions can be found in \cite{Hopfmuller2017a,
Hopfmuller2018}.  The boundary 
term $\ell$ that arises in the decomposition is constructed from the 
inaffinity $k$ of the null generator $l^a$, and has appeared 
in previous analyses on null boundary terms in the action for 
general relativity \cite{Parattu_2016, Hopfmuller2017a,Lehner_2016}.
In particular, we find additional flux terms beyond those 
employed in \cite{Haco2018, Chen:2020nyh}, whose presence is 
necessary to ensure that the flux is independent of the choice 
of auxiliary null vector $n_a$.

With all this in place, we give a systematic analysis in section 
\ref{killinghorizon}
of the quasilocal charges in the HHPS construction, as well as 
the generalization to arbitrary bifurcate, axisymetric
Killing horizons \cite{Haco2018, Chen:2020nyh}. The symmetry
algebra consists of two copies of the Virasoro algebra,
and the central charges are computed to be
\beq
c = \bar c = \frac{3A}{\pi G (\alpha+\bar \alpha)}, 
\eeq
where $\alpha$ and $\bar\alpha$ are two parameters characterizing
the symmetry generators, and are related to the choice of 
left and right temperatures.  These values of $c$, $\bar c$ are 
twice the value given in \cite{Haco2018, Chen:2020nyh}, and consequently, when 
applying the Cardy formula in section \ref{sec:canonCardy}, we find that the entropy is twice the 
Bekenstein-Hawking entropy of the horizon. We take this as an 
indication that the quasilocal charge algebra is sensitive to degrees
of freedom associated with the complementary region. In particular,
we note that the factor of $2$ could be explained if 
the central charge appearing in the Barnich-Troessaert bracket
was associated with a pair of quasilocal charge algebras, one 
on each side of the dividing surface.  This interpretation is further
motivated by the conjectured 
edge mode contribution to entanglement entropy in 
gravitational theories, which employ such a pair of quasilocal 
charges at an entangling surface \cite{Donnelly2016a}. The doubling 
of $c$, $\bar c$ would then be intimately related to the fact 
that we are considering an open system that is interacting 
with its complement. Conversely, if the charges were instead 
integrable so 
that they lived in a closed system, we would expect the standard
entropy to arise via the Cardy formula. We demonstrate that this 
is the case in sections \ref{sec:integrable} and 
\ref{sec:microcanonical}  by showing that 
a different boundary term is needed in order to find integrable
generators.  The new boundary term halves the value of the central
charges and the entropy, and also leads to agreement between
the microcanonical and canonical Cardy formulas.

In section \ref{sec:discussion}, we further discuss the interpretation of these 
results, and describe some directions for future work.  

\paragraph{Note added:} This work is being released in coordination
with \cite{Chen2020}, which explores some related topics.

\subsection{Notation}
We work in arbitrary spacetime dimension $d$ with metric
signature $(-,+,+,\ldots)$. 
Spacetime tensors
will be written with abstract indices $a, b, \ldots $, such as the metric $g_{ab}$. We denote null hypersurfaces by $\NNN$, and indices $i,j,\ldots$ will denote tensors defined on $\NNN$, such as $q_{ij}$ and $l^k$. An equality that only holds at the location of 
$\NNN$ in spacetime will be written as $\hateq$. 
Differential forms will often be written without indices, and, when
necessary, we distinguish a form $\theta$ defined on spacetime from
its pullback $\p\theta$ to $\NNN$ using boldface. 
The null normal to $\NNN$ will be denoted $l_a$, and the auxiliary null vector will be denoted $n^a$. The volume form on spacetime is denoted $\ep$, and occasionally it will be written as $\ep_a$ or $\ep_{ab}$ when the displayed indices are being contracted; the undisplayed indices are left implicit. The volume form on $\NNN$ induced from $l_a$ will be denoted $\eta$, and the horizontal spatial volume form on $\NNN$ will be denoted $\mu$. The notation for the contraction of a vector $v^a$ into
a differential form $m$ is $i_v m$. The notation for operations defined on $\SSS$, the space of solutions 
to the field equations, is described in section \ref{covphase} below,
including definitions of $I_{\hat\xi}$, $L_{\hat\xi}$, $\delta$,
and $\Delta_{\hat\xi}$.

\section{Quasilocal charge algebra}
We begin by reviewing the covariant phase space construction
in section \ref{covphase}, before
turning to the construction of quasilocal charges in 
section \ref{sec:quasicharge}, and 
their algebra in section \ref{sec:BTbrack}. Section
\ref{sec:quasicharge} explains the relation between the 
Wald-Zoupas construction \cite{Wald2000b} and the recent work
by Harlow and Wu on the covariant phase space with boundaries 
\cite{Harlow2019}.  This yields an unambiguous definition
of the quasilocal charges by the arguments of \cite{Harlow2019},
once the form of the flux $\beom$ has been specified.  To
fix this final ambiguity, we require that the flux be of 
Dirichlet form, and we discuss the motivation for this 
choice coming from the combined variational principle for the
subregion and its complement.  The algebra of charges
is then defined in section \ref{sec:BTbrack}, where we 
give a general expression for the extension of the algebra
in terms of the anomaly of the boundary term appearing 
in the symplectic potential decomposition.  

\subsection{Covariant phase space}\label{covphase}
The main tool we employ in constructing the quasilocal
charge algebra is the covariant phase space
\cite{Witten1986, Crnkovic1987, Crnkovic:1987tz, Ashtekar1991, Lee1990a}.\footnote{We
largely follow the notation of \cite{Speranza2018a}
when working with the covariant phase space.} It provides
a canonical description of field theories without
singling out a preferred time foliation, and therefore is 
well-suited for handling diffeomorphism-invariant theories,
such as general relativity.  Covariance is achieved by working 
with the space $\mathcal{S}$ of solutions to the field equations,
as opposed to the space of initial data on a time slice.  

$\mathcal{S}$ 
can be viewed as an infinite-dimensional manifold, 
on which many standard differential-geometric techniques apply.  
Fields such as the metric $g_{ab}$ can be viewed as functions on $\SSS$, 
and their variations, such as $\delta g_{ab}$, are one-forms.  The operation $\delta$ 
of taking variations can be viewed as the exterior derivative on $\SSS$, and 
forms of higher degree can be built by taking exterior derivatives and 
wedge products in the usual way.  The product of two differential forms 
$\alpha$ and $\beta$ on $\SSS$ will always implicitly be a wedge product, so
that $\alpha\beta = (-1)^{\deg(\alpha)\deg(\beta)} \beta\alpha$,
which allows the symbol $\wedge$ to exclusively 
denote the wedge product between 
differential forms on the spacetime manifold $\mathcal{M}$.  We 
denote by $I_V$ the operation of contracting a vector field $V$ on $\SSS$
with a differential form.  Functions of the form $h_{ab} = I_V\delta g_{ab}$
are simply solutions to the linearized field equations, and so the 
vector fields on $\SSS$ are seen to coincide with the space of linearized 
solutions.  

Since diffeomorphisms of $\mathcal{M}$ are gauge symmetries of general
relativity, they define an important subclass of linearized solutions 
$h_{ab} = \lie_\xi g_{ab}$, where $\xi^a$ is a spacetime vector field.  
The corresponding vector field on $\SSS$ generating this transformation
will be called $\hat\xi$, which satisfies 
$I_{\hat\xi}\delta g_{ab} = \lie_\xi g_{ab}$.  
Note also that $I_{\hat{\xi}}\delta g_{ab} = L_{\hat\xi} g_{ab}$, 
where $L_{\hat\xi}$ is the Lie derivative along the vector $\hat\xi$ 
in $\SSS$, and hence $L_{\hat\xi}$ and $\lie_\xi$ agree when acting 
on the metric $g_{ab}$.  The action of $L_{\hat\xi}$ on higher order 
differential forms on $\SSS$ can be computed via the Cartan
formula $L_{\hat\xi} = I_{\hat\xi}\delta + \delta I_{\hat\xi}$.  Any
differential form $\alpha$ that is locally constructed from 
dynamical fields and for which $L_{\hat\xi}\alpha = \lie_\xi\alpha$
will be called {\it covariant} with respect to $\hat\xi$. Since we later work with noncovariant objects as well, it is useful to 
define the anomaly operator 
\beq\label{eqn:anomop}
\Delta_{\hat\xi} = L_{\hat\xi}-\lie_\xi, 
\eeq
as in
\cite{Hopfmuller2018}, which measures 
the failure of a local object to be covariant.  We therefore
also refer to $\Delta_{\hat\xi} \alpha$ as the {\it noncovariance}
or {\it anomaly}
of $\alpha$ with respect to $\hat \xi$.  
As we will see, $\Delta_{\hat\xi}$ plays a prominent
role in characterizing the extensions that appear in 
quasilocal charge algebras, and the anomalies it computes are, 
in many ways,  classical analogs of the anomalies that 
appear in quantum field theories.  In particular, as we show in 
appendix \ref{app:anom}, $\Delta_{\hat\xi}$ satisfies
\beq \label{eqn:WZcons}
[\Delta_{\hat\xi}, \Delta_{\hat \zeta}] = -\Delta_{\widehat{[\xi,\zeta]}},
\eeq
which, when imposed on the functionals of the theory, is the direct analog of the Wess-Zumino consistency
condition for quantum anomalies \cite{Wess1971}.\footnote{
See \cite{Shyam2018} for a discussion of the Wess-Zumino consistency
condition in the context of holographic Weyl anomalies.}

The covariant phase space arises from $\SSS$ by imbuing it with
a presymplectic form.  To construct it, one begins
with the Lagrangian of the theory, $L$, which is a spacetime top form
whose variation satisfies
\beq
\delta L = E^{ab}\delta g_{ab} + d\theta,
\eeq
where $E^{ab}=0$ are the classical field equations, and $\theta$ is a one-form
on $\SSS$ and a $(d-1)$-form on spacetime called the symplectic 
potential current.  For general relativity, the various quantities are 
\begin{align}
L &= \frac1{16\pi G} (R-2\Lambda) \ep \\
E^{ab} &= \frac{-\ep}{16\pi G} \left(R^{ab}-\frac12 R g^{ab}+\Lambda g^{ab}\right)\\
\theta &= \frac{1}{16\pi G} \ep_a\Big(g^{bc}\delta\Gamma^a_{bc}
-g^{ac}\delta\Gamma^b_{bc}\Big),
\end{align}
where the variation of the Christoffel symbol is
\beq
\delta\Gamma^a_{bc} = \frac12 g^{ad}\left(\nabla_b\delta g_{dc}
+\nabla_c\delta g_{bc} - \nabla_d\delta g_{bc}\right),
\eeq
and we recall that $\ep_a$ still denotes the spacetime volume
form, with uncontracted indices not displayed.  

The $\SSS$-exterior derivative of $\theta$ defines the symplectic current
$\omega = \delta \theta$, and its integral over a Cauchy surface $\Sigma$ for
the region of spacetime under consideration yields the presymplectic form,
\beq
\Omega = \int_\Sigma \omega.
\eeq
$\Omega$ is called ``presymplectic'' because it contains
degenerate directions corresponding to diffeomorphisms of $\MMM$.  Since
diffeomorphisms are symmetries of the Lagrangian, they lead to Noether
currents that are conserved on shell, given by
\beq\label{eqn:Jxi}
J_\xi = I_{\hat\xi} \theta - i_\xi L.
\eeq
Because $dJ_\xi = 0$ identically for all vectors $\xi^a$, the Noether
current can be written as the exterior derivative of a potential,
$J_\xi = d Q_\xi$, which is locally constructed from the metric; 
for general relativity, 
this potential is \cite{Wald1990a, Wald1993a},
\beq
Q_\xi = \frac{-1}{16\pi G}\ep\indices{^a_b}\nabla\ind{_a} \xi^b.
\eeq
The degeneracy of $\Omega$ follows straightforwardly from computing the 
contraction with $I_{\hat\xi}$, 
\beq \label{eqn:IxiOm}
-I_{\hat\xi}\Omega = \int_{\partial\Sigma} \Big(\delta Q_\xi - i_\xi \theta\Big),
\eeq
using the fact that $\theta$ is 
covariant, $I_{\hat\xi}\delta \theta + \delta I_{\hat\xi}\theta = \lie_\xi \theta$ \cite{Iyer1994a}.  Since this contraction localizes
to a boundary integral, any diffeomorphism that acts purely in the interior
is a degenerate direction of $\Omega$.  The phase space $\mathcal{P}$ 
is a quotient of $\SSS$ by the degenerate directions, onto which $\Omega$ 
descends to a nondegenerate symplectic form \cite{Lee1990a}.

\subsection{Quasilocal charges}\label{sec:quasicharge}
According to (\ref{eqn:IxiOm}), 
diffeomorphisms with support near the Cauchy surface boundary
$\partial\Sigma$ are not degenerate directions; rather, they lead to 
a notion of quasilocal charges associated with the subregion defined by 
$\Sigma$.  In the case that $\xi^a$ at $\partial\Sigma$ is vanishing or 
tangential, the term $i_\xi\theta$ in (\ref{eqn:IxiOm}) drops out 
when pulled back to $\partial\Sigma$, and a Hamiltonian for the 
transformation can be defined by
\beq
H_\xi = \int_{\partial\Sigma} Q_\xi,
\eeq
which generates the symmetry transformation on phase space via Hamilton's
equations,
\beq\label{eqn:Ham}
\delta H_\xi = -I_{\hat\xi}\Omega.
\eeq

When $\xi^a$ is not tangential to $\partial\Sigma$,  $-I_{\hat\xi}\Omega$
generally cannot be written as a total variation, unless boundary 
conditions are imposed so that 
$\int_{\partial\Sigma} i_\xi\theta = \delta B_\xi$ for some quantity $B_\xi$. 
Such boundary conditions are natural when $\partial\Sigma$ sits at an 
asymptotic boundary, but not at boundaries associated with 
subregions of a larger system, where the boundary conditions are
generically inconsistent with the global dynamics.  Instead, one can 
define a quasilocal charge associated with the transformation
following the Wald-Zoupas prescription \cite{Wald2000b}.  The
quasilocal charge
is not conserved since it fails to satisfy Hamilton's equation
(\ref{eqn:Ham}), but it satisfies a modified equation that 
relates the nonconservation to a well-defined flux through the boundary
of the subregion.

\begin{figure}[t]
\centering
\includegraphics[width = 0.5\textwidth]{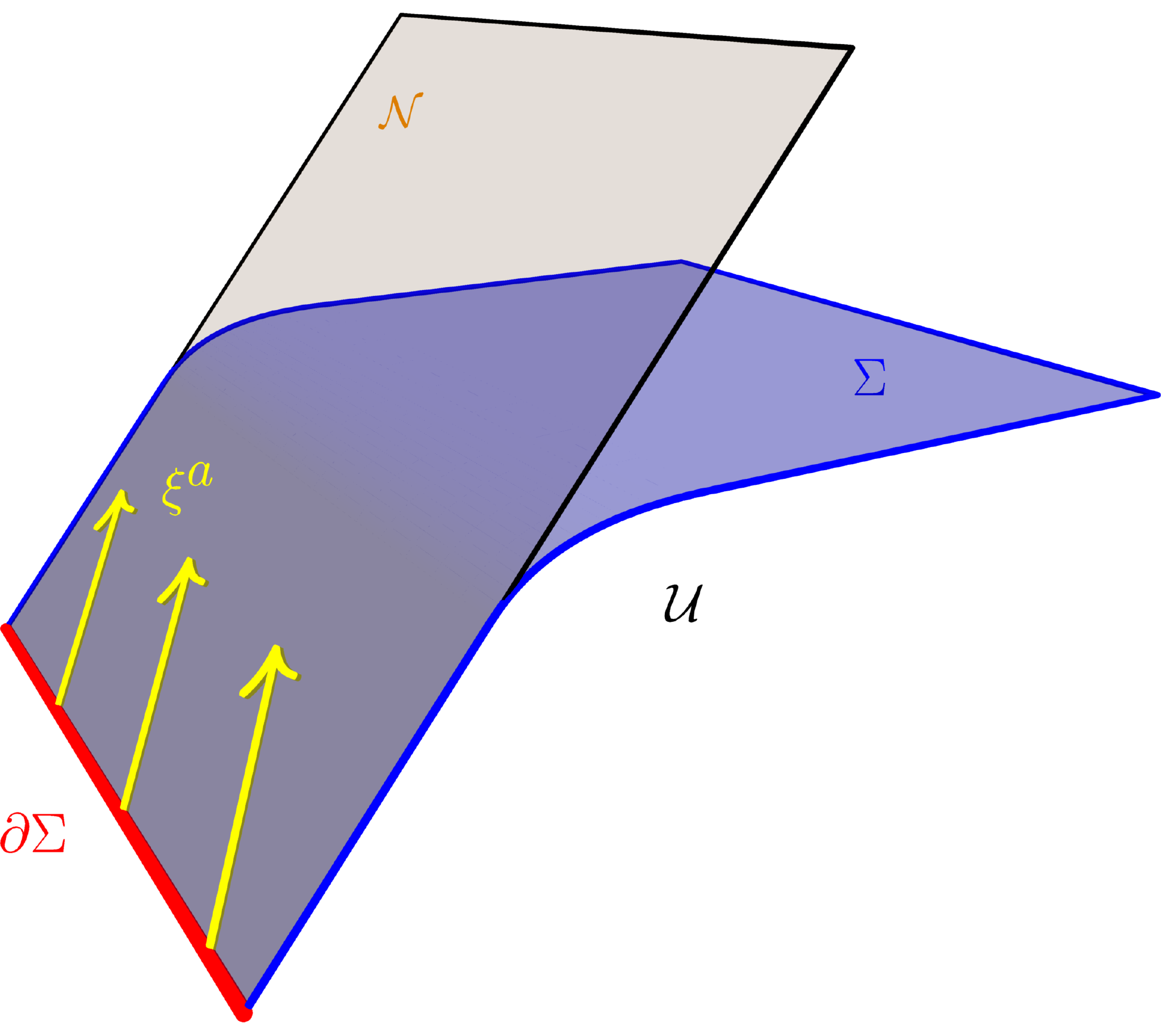}
\caption{In the Wald-Zoupas construction, one 
seeks to construct quasilocal charges for a 
transformation generated by $\xi^a$, which is 
tangent to a hypersurface $\NNN$ bounding
an open subregion $\mathcal{U}$ to the right of $\NNN$.  
The charges are constructed
as integrals over a codimension-2 surface $\partial\Sigma$, bounding a Cauchy surface $\Sigma$
for the subregion.  The vector field
$\xi^a$ can have both tangential and normal components 
to $\partial\Sigma$.  In this figure, $\NNN$ is 
a null hypersurface, and the Cauchy surface has been chosen
to include a segment of $\NNN$.}
\label{fig:wz}
\end{figure}

Here, we give a presentation of the Wald-Zoupas construction,
using the formalism developed by Harlow and Wu \cite{Harlow2019}
for dealing with
boundaries in the covariant phase space.\footnote{See also 
\cite{Shi2020}
for a similar recent application of Harlow and Wu's formalism
to the Wald-Zoupas construction. }  
The Wald-Zoupas construction begins with a subregion of spacetime
$\sub$, bounded 
by a hypersurface $\NNN =\partial\sub$ (see figure
\ref{fig:wz}).  
Later  $\NNN$ will be taken to be a null
hypersurface, but the present discussion applies more 
generally for any signature of $\NNN$.  
On $\NNN$, one looks for a decomposition of the pullback $\p\theta$ 
of the symplectic potential of the 
following form 
\beq\label{eqn:thdecomp}
\p\theta = -\delta  \ell +d  \beta +  \beom
\eeq
where $\ell$ is referred to as the {\it boundary term}, $\beta$ is the 
{\it corner term}, and $\beom$ is the {\it flux term}. 
The reason for this terminology becomes apparent
from the variational principle for the theory defined in the subregion $\sub$ \cite{Harlow2019,Iyer1995c}.
The action for the subregion is 
\beq
S = \int_{\sub} L + \int_{\NNN}  \ell,
\eeq
and by the decomposition (\ref{eqn:thdecomp}) the variation satisfies
\beq
\delta S = \int_\sub E^{ab}\delta g_{ab} + \int_\NNN \left(\beom + d\beta\right),
\eeq
and so the action is stationary when the bulk field equations $E^{ab}=0$
hold and boundary conditions are chosen to make $ \beom$ vanish, with the $d\beta$ term localizing to the boundary
of $\NNN$, i.e.\ the corner. 
In the Wald-Zoupas setup, 
boundary conditions to make $\beom$ vanish are not imposed;
instead, $\beom$ is used to construct the fluxes of the quasilocal charges.  In \cite{Wald2000b}, the combination $\beom + d\beta$ is referred to as a 
potential for the pullback of $\omega$ to $\NNN$, since 
by equation (\ref{eqn:thdecomp}) we see that\footnote{In \cite{Wald2000b} the combination $\mathcal{E} + d\beta $ was denoted $\Theta$.} \begin{align}\delta (\beom + d\beta)
 = \delta \p\theta = \p\omega. 
 \end{align} 

The corner term $\beta$ is used to modify the symplectic form for the 
subregion.\footnote{This type of modification, for example, gives the difference
between the covariant 
Iyer-Wald symplectic form and the standard ADM symplectic form,
see \cite{Burnett1990}, and also recent discussions of this point 
in \cite{Harlow2019, Freidel2020}.}  This is done by extending $d\beta$ to an exact form on all of 
$\sub$, and then treating $\theta - d\beta$ as the 
symplectic potential current.  The symplectic form then becomes
\beq
\Omega = \int_\Sigma \omega - \int_{\partial\Sigma} \delta \beta.
\eeq
We can then evaluate the contraction of $\Omega$ with a diffeomorphism 
generator $\xi^a$ that is parallel to $\NNN$, but not
necessarily to $\partial\Sigma$,
\begin{align}
-I_{\hat\xi} \Omega &= \int_{\partial\Sigma} \Big( \delta Q_\xi - i_\xi\theta
+I_{\hat\xi}\delta\beta\Big) \nonumber \\
&= \int_{\partial\Sigma} \Big( \delta Q_\xi + i_\xi\delta\ell 
-\delta I_{\hat\xi} \beta\Big)
 - \int_{\partial\Sigma} \Big( i_\xi \beom
 -\Delta_{\hat\xi}\beta\Big). \label{eqn:dH+flux}
\end{align}
The first term is the total variation of a quantity 
\beq \label{eqn:Hxi}
H_\xi = \int_{\partial\Sigma}\Big(Q_\xi + i_\xi\ell -I_{\hat\xi}\beta\Big),
\eeq
which
we call the quasilocal charge for the transformation.  The second
term in (\ref{eqn:dH+flux}) represents the failure of the quasilocal
charge to be an integrable generator of the symmetry.  Assuming that
$\beta$ is covariant, so that $\Delta_{\hat\xi}\beta =0$, 
the obstruction to integrability of the charge is simply given
by the integral of the flux density $i_\xi \beom$.  With slight
modifications, the case where $\Delta_{\hat\xi}\beta\neq 0$ can
be handled, and is described in appendix \ref{app:cornerimprove}.  Equation
(\ref{eqn:dH+flux}) can be rearranged slightly to take the form of a
modified Hamilton's equation,
\beq\label{eqn:Hamflux}
\delta H_\xi = -I_{\hat\xi}\Omega + \int_{\partial\Sigma} i_\xi \beom
\eeq

To further the interpretation of $\beom$ as a flux
of $H_\xi$, we note 
first that the integrand of (\ref{eqn:Hxi}) is defined on all
of $\NNN$, and its exterior derivative can be computed as 
\begin{align}
d\left(Q_\xi +i_\xi \ell -I_{\hat\xi}\beta\right)
&=
I_{\hat\xi}\p\theta - i_\xi L -i_\xi d\ell 
+\lie_\xi\ell -I_{\hat\xi}d\beta \nonumber\\
&=
I_{\hat\xi}\beom -\Delta_{\hat\xi}\ell -i_\xi(L + d\ell)
\label{eqn:fluxdensity}
\end{align}
Integrating this relation on a segment $\NNN_1^2$ 
of $\NNN$ between two cuts $S_2$ and $S_1$,
and using that $\xi^a$ is parallel to $\NNN$ yields
\beq\label{eqn:continuity}
H_\xi(S_2)-H_\xi(S_1) = 
\int_{\NNN_1^2} \Big(I_{\hat\xi}\beom -\Delta_{\hat\xi} \ell\Big).
\eeq
This can be interpreted as an anomalous continuity equation for the quasilocal
charge $H_\xi$: the difference in the charge between two cuts 
is simply given by the flux $F_\xi = 
\int_{\NNN_1^2} I_{\hat\xi}\beom$, up to an
anomalous contribution from $\Delta_{\hat\xi}\ell$.  
This anomalous term in the flux vanishes if $\ell$ is covariant
with respect to $\xi^a$; however, we will find 
that on null surfaces,
the most natural choice for the flux term $\beom$ requires a boundary
term that is not covariant.  Note that this equation differs from the standard continuity equation 
derived in the Wald-Zoupas and related constructions
\cite{ CFP,Wald2000b, Shi2020,Adami2020a}, 
which assume a covariant boundary term, so that  $\Delta_{\hat\xi}\ell$ drops out.  This is the 
first indication that the noncovariance of the boundary term can be interpreted as an anomaly, since it behaves as an explicit violation of a contintuity equation for the quasilocal charges. 
In quantum field theory, anomalies play a similar role to that of
$\Delta_{\hat\xi} \ell$, where they lead to explicit violations 
of the Ward identities.

Up to this point, we have placed no restrictions on the precise
form of the flux $\beom$.  Equation (\ref{eqn:thdecomp}) does
not uniquely specify $\beom$, since it can always be shifted 
by terms of the form $\beom \rightarrow \beom - \delta b - d\lambda$
by making compensating changes $\ell \rightarrow \ell - b$, $\beta
\rightarrow \beta + \lambda$.  These ambiguities in $\beom$ 
are similar in appearance to the standard  
Jacobson-Kang-Myers ambiguities \cite{Jacobson1994b, Iyer1994a} 
in the definition of the 
symplectic potential current, in which $\theta\rightarrow
\theta + \delta b' + d\lambda'$.  Although the $(b,\lambda)$ and 
$(b',\lambda')$ ambiguities are in principle distinct, they 
can  be used in tandem to leave
$\beom$ invariant, by setting $(b,\lambda) = (b',\lambda')$.  Additionally, 
the charge densities $h_\xi = Q_\xi + i_\xi \ell - I_{\hat\xi}\beta$
are also unchanged, provided one shifts the Noether potential by
$Q_\xi + i_\xi b' + I_{\hat\xi} \lambda'$, as was recently 
emphasized by \cite{Harlow2019}.  These transformations
of $Q_\xi$ simply follow from its definition as a potential
for the Noether current $J_\xi$ (\ref{eqn:Jxi}) as long as
one assumes that $b'$ is covariant (no assumption on the covariance
properties of $\gamma'$ is needed).  

Thus, in order to avoid the ambiguities just described, we
need to fix the form of the flux  $\beom$.  As discussed in
\cite{Iyer1995c, Bart2019}, different choices for $\beom$ 
are related to different boundary conditions one would 
impose to make the flux vanish.  
The principle we will advocate for in this work is that 
the flux take a Dirichlet form, which,\footnote{This coincides with 
the ``canonical boundary conditions'' discussed in \cite{Bart2019}.}
for $\NNN$ timelike or spacelike,  means it is written as 
\beq
\beom = \pi^{ij} \delta g_{ij}, \label{fluxcond}
\eeq
where $\delta g_{ij}$ is the metric variation pulled back to 
$\NNN$, constituting the intrinsic data on the surface, and $\pi^{ij}$ is a symmetric-tensor-valued top form on 
$\NNN$ constructed from the extrinsic data, and interpreted as the conjugate momenta to $\delta g_{ij}$.  The intrinsic data on a 
null surface is slightly different since the induced metric is 
degenerate, and so it is taken to also include variations 
of the null generator $\delta l^i$, leading to the null Dirichlet
flux condition
\beq\label{eqn:fluxcondnull}
\beom = \pi^{ij}\delta g_{ij} + \pi_i \delta l^i.
\eeq
Dependence on non-intrinsic components of the metric, such as the 
lapse and shift, is removed by the choice of corner term, which
further fixes the ambiguities in specifying the flux.  
Imposing the Dirichlet form on $\beom$ greatly reduces the freedom
in its definition, since most of the ambiguities will
involve variations of quantities constructed from the extrinsic geometry of 
$\NNN$.  We will find that for general relativity, the Dirichlet
requirement fixes $\beom$ essentially uniquely.\footnote{For 
asymptotic symmetries, it can be important to include objects 
constructed from the intrinsic curvature of the metric, in
order to have finite symplectic fluxes at infinity, 
which then modifies
$\pi^{ij}$ when imposing the Dirichlet form
\cite{Henningson1998a, Balasubramanian1999a, 
DeHaro2001, Papadimitriou2005, Mann2006, Compere2011,
Compere2018, Compere2019b}.  Such terms will not 
be important for our analysis of a null boundary at a finite
location. }  

One reason for favoring the Dirichlet form of the flux comes from
considering the variational principle for a subregion $\sub$ 
and its complement $\bar\sub$.  When gluing the subregions
across the boundaries $\NNN$ and $\bar\NNN$, 
the Dirichlet form of $\beom$ is used when
kinematically matching the intrinsic quantities on $\NNN$.
Viewed from one side, this takes the form of a Dirichlet
condition, with the value of $g_{ij}$ on one side fixed
by the value on the other side.  
Upon identifying $\NNN$ with $\bar\NNN$, 
matching $g_{ij}$, and imposing the bulk field equations, the 
variation of the action is given by
\begin{align}
\delta\left(\int_\sub L + \int_\NNN \ell + \int_{\bar\NNN} \bar\ell
+ \int_{\bar\sub} L \right)
= \int_{\NNN}(\pi^{ij}-\bar\pi^{ij})\delta g_{ij} +\text{corner term}.
\end{align}
Stationarity of the action then dynamically sets $\pi^{ij} -
\bar\pi^{ij} = 0$, or more generally equal to the distributional
stress energy on $\NNN$ if present, according to the 
 junction
conditions \cite{PhysRevD.43.1129, poisson2004a}.  
If instead a Neumann form for the flux
$\beom^N = -g_{ij}\delta\pi^{ij}$ were employed, the matching
condition would kinematically set $\pi^{ij} = \bar\pi^{ij}$,
and then $g_{ij}-\bar{g}_{ij}$ would dynamically be set to 
zero.  In this case, there does not appear to be a straightforward
way to allow for distributional stress-energy on $\NNN$. 
In vacuum, the end result is classically the same, with
both $g_{ij}$ and $\pi^{ij}$ matching at $\NNN$, although
already the Dirichlet form has the advantage of allowing for the
presence of distributional stress-energy.  
In a quantum description, these two options differ even more.
Since the path integral receives contributions from off-shell
configurations, 
the Dirichlet
matching appears to be preferred, since the Neumann matching
allows for discontinuities in the intrinsic metric, which produce
distributionally ill-defined curvatures \cite{poisson2004a}.\footnote{These singularities
are unlike conical defects, whose curvature is well-defined 
as a distribution and are therefore valid configurations in the 
path integral.}
We further discuss the Dirichlet matching condition in
section \ref{BTdisc}.

\subsection{Barnich-Troessaert bracket} \label{sec:BTbrack}

Having defined the quasilocal charges $H_\xi$ given
by (\ref{eqn:Hxi}) for the diffeomorphisms 
generated by $\xi^a$, we now consider the problem of computing their algebra. 
In standard Hamiltonian mechanics, this 
is given by the Poisson bracket constructed from the symplectic
form of the system.  When the charges are integrable, so that they 
satisfy Hamilton's equation (\ref{eqn:Ham}), the Poisson bracket
can be evaluated by contracting the vector fields generating the 
symmetry into the symplectic form,
\beq \label{eqn:Poissonbracket}
\{ H_\xi , H_\zeta \} = -I_{\hat\xi} I_{\hat\zeta}\Omega = 
-\left(H_{[\xi,\zeta]} + K_{\xi,\zeta}\right).
\eeq
The second equality in this equation is a statement of the fact
that Poisson brackets must reproduce the Lie bracket of the 
vector fields $\xi^a$, $\zeta^a$, up to a central extension, 
denoted $K_{\xi,\zeta}$.\footnote{There are two related
reasons for the minus sign appearing in (\ref{eqn:Poissonbracket}). 
The first is that the Poisson bracket reproduces the Lie
bracket $[\hat\xi,\hat\zeta]_\SSS$ of vector fields on $\SSS$, 
which, as shown in (\ref{eqn:SLie}), 
is minus the spacetime Lie bracket
for field-independent vector fields.  It arises because 
diffeomorphisms give a left action on spacetime, but a 
right action on $\SSS$.  The second reason
is that the Hamiltonians are representing the Lie algebra
of the diffeomorphism group, whose Lie bracket is minus the 
vector field Lie bracket \cite{Milnor1984}.  }

For quasilocal charges, 
their failure to satisfy Hamilton's equations due to the flux
term in (\ref{eqn:Hamflux}) prevents a naive application 
of (\ref{eqn:Poissonbracket}) to their brackets.  Instead, Barnich
and Troessaert \cite{Barnich:2011mi} proposed a modification to the bracket that 
accounts for the nonconservation of the charges due to the loss
of flux from the subregion.  When the corner term $\beta$ is 
covariant, their bracket is given by
\beq \label{eqn:BTbracket}
\{H_\xi , H_\zeta\} = -I_{\hat\xi} I_{\hat\zeta}\Omega +\int_{\partial\Sigma}
\Big(i_\zeta I_{\hat\xi}\beom - i_\xi I_{\hat\zeta} \beom\Big),
\eeq
where we see that the bracket is modified by the fluxes 
$F_\xi =\int_{\partial\Sigma} I_{\hat\xi}\beom$ 
identified in the Wald-Zoupas construction.  
A heuristic way to understand this equation is as follows: imagine adding an
auxiliary system which collects the flux lost through $\NNN$ when
evolving along $\xi^a$ (for example, this could just be the phase
space associated with the complementary region $\bar\sub$).  
The total system consisting of the subregion
and the auxiliary system is assumed to have a Poisson bracket defined on it,
such that $\hat\xi$ is a symmetry of the bracket in the usual sense.  
The Hamiltonian for $\hat\xi$ should be a sum of the quasilocal Hamiltonian
$H_\xi$ and a term $H_\xi^\text{aux}$ associated with the auxiliary system.  
Hamilton's equation for the total system then reads
\beq \label{eqn:H+aux}
I_{\hat\xi}\delta H_\zeta = \{H_\xi + H_\xi^\text{aux}, H_\zeta\}.
\eeq
The contribution from $\{H_\xi^\text{aux}, H_\zeta\}$ should compute
the flux of $H_\zeta$ into the auxiliary system due to an infinitesimal
change of $\partial\Sigma$ along $\xi^a$, which is just the 
integral of $i_\xi I_{\hat\zeta} \beom$, given our identification of 
$I_{\hat\zeta} \beom$ with the flux density.  
Equation (\ref{eqn:H+aux}) then becomes
\beq
I_{\hat\xi}\delta H_\zeta = \{H_\xi, H_\zeta\} 
+ \int_{\partial\Sigma} i_\xi I_{\hat\zeta}\beom,
\eeq
which  reduces to (\ref{eqn:BTbracket}) after using the expression
(\ref{eqn:Hamflux}) for $\delta H_\zeta$.  Going forward, we will
take (\ref{eqn:BTbracket}) as the definition of the bracket for the 
quasilocal charges, and delay further discussion of its interpretation
to section \ref{BTdisc}.  

An important property of the Barnich-Troessaert bracket is that 
it reproduces the Lie bracket algebra of the vector fields, up to
abelian extensions \cite{Barnich:2011mi, Troessaert:2015nia}. 
This can be explicitly
verified using the expression (\ref{eqn:Hxi}) for the quasilocal
charges, and an exact expression for the extension $K_{\xi,\zeta}$ 
can be given.  After a short calculation (see appendix \ref{app:BTident}), 
one finds 
\begin{align}
\{H_\xi, H_\zeta\} &= -\left(H_{[\xi,\zeta]} +K_{\xi,\zeta} \right) \label{eqn:HxiHzeta}\\
K_{\xi,\zeta} &= \int_{\partial\Sigma} \Big( i_\xi\Delta_{\hat\zeta}\ell - 
i_\zeta \Delta_{\hat\xi} \ell \Big).\label{eqn:Kxizeta}
\end{align}
Hence, we arrive at one of the main results of this work, namely, that
the extension $K_{\xi,\zeta}$ is determined entirely by the noncovariance
of the boundary term, $\Delta_{\hat\xi} \ell$.  As an immediate
corollary, we see that 
the extension $K_{\xi,\zeta}$ always vanishes if the boundary
term $\ell$ is covariant with respect to the generators $\xi^a$. 
Equation (\ref{eqn:Kxizeta}) remains valid even when 
boundary conditions are imposed to ensure the transformation has 
integrable generators.  In this case, the fluxes in (\ref{eqn:BTbracket}) vanish, and we see that the Barnich-Troessaert
bracket reduces to a Dirac bracket on the subspace of field
configurations that satisfy the boundary conditions.  This 
therefore gives a universal formula for the central extension
in these cases, in addition to the more general cases involving
nonintegrable generators. 

It is worth emphasizing that the central charge appears in this formula because 
we have chosen to fix a background structure in defining the boundary, 
which gives rise to nonzero anomalies $\Delta_{\hat\xi} \ell$.  However,  
the value of $K_{\xi,\zeta}$ does not depend on the choice 
of constant added to the Hamiltonians, which, for example, 
could be chosen to ensure that the
Hamiltonians vanish in a given background solution.  More precisely,
different choices for these 
constant shifts can only change the extension by trivial constant terms of the 
form $C_{[\xi,\zeta]}$, which will not 
change the 2-cocycle that  $K_{\xi,\zeta}$ represents for the 
Lie algebra of the vector fields $\xi^a, \zeta^a$.   In particular, $C_{[\xi,\zeta]}$ cannot be chosen to cancel $K_{\xi,\zeta}$ if the extension
comes from a nontrivial 2-cocycle, as occurs in the Virasoro example
we consider in section \ref{killinghorizon}.  

In general, the new generators $K_{\xi,\zeta}$ are not central,
since they are allowed to transform nontrivially under the action of 
another generator $H_\chi$.  Instead, they give an abelian
extension of the algebra by defining their brackets to be 
\begin{align}
\{ H_\chi, K_{\xi,\zeta}\} &= I_{\hat\chi} \delta K_{\xi,\zeta}\label{eqn:HK} \\
\{K_{\xi,\zeta},K_{\chi,\psi}\} &=0.
\end{align}
This algebra closes provided  $I_{\hat\chi} \delta K_{\xi,\zeta}$ 
is expressible as a sum of other generators $K_{\xi',\zeta'}$, and 
the Jacobi identity holds as long as $K_{\xi,\zeta}$ satisfies a generalized
cocycle condition \cite{Barnich:2011mi},
\beq \label{eqn:Kcocycle}
I_{\hat\chi} \delta K_{\xi,\zeta} + K_{[\chi,\xi],\zeta} + 
(\text{cyclic $\chi\rightarrow\xi\rightarrow\zeta$}) =0.
\eeq
Of course, when the right hand side of (\ref{eqn:HK}) vanishes,
$K_{\xi,\zeta}$ represents a central extension of the algebra.  

We verify the above cocycle condition for (\ref{eqn:Kxizeta}) in appendix
\ref{app:BTident}. We should expect this to be the case because 
$K_{\xi,\zeta}$ in (\ref{eqn:Kxizeta}) is of the form of a trivial
field-dependent 2-cocycle, 
in the terminology of \cite{Barnich:2011mi}.\footnote{For an
interpretation of this field-dependent extension
in terms of a Lie algebroid in the example of $\text{BMS}_4$
asymptotic symmetries, see \cite{Barnich2017}.}
That is, it can be expressed as 
\beq\label{eqn:trivcocycle}
K_{\xi,\zeta} = I_{\hat\zeta} \delta B_\xi - I_{\hat\xi} \delta B_\zeta -
B_{[\xi,\zeta]},  \qquad B_\xi\equiv \int_{\partial\Sigma} i_\xi \ell
\eeq
Despite this terminology, $K_{\xi,\zeta}$ is certainly not required to be 
trivial as a cocycle for the Lie algebra generated by the vector fields. 
This will be explicitly demonstrated for the algebra considered in 
section \ref{killinghorizon}, in which case $K_{\xi,\zeta}$ becomes the nontrivial 
central extension of the Witt algebra to Virasoro.  

Finally, it is worth noting that the corner term $\beta$, although
important in arriving at the Dirichlet form  (\ref{fluxcond}) or
(\ref{eqn:fluxcondnull}) for the flux, is not important for obtaining
the 
correct algebra for the quasilocal charges, including the 
extension $K_{\xi,\zeta}$.  Algebraically, the $\beta$ term
in the quasilocal charge is functioning as a trivial extension of the algebra, since the $\beta$ terms do not mix with other terms 
when deriving the identity (\ref{eqn:HxiHzeta}), as discussed in 
appendix \ref{app:BTident}.  This is the reason that the central
charges computed in \cite{Haco2018, Chen:2020nyh} were 
correctly identified, even without taking corner terms into
account.

\section{Symplectic potential on a null boundary} \label{sec:symppot}

In this section, we apply the covariant phase space formalism to null boundaries. We decompose the symplectic potential into boundary, corner, and flux terms, and describe the resulting canonical pairs on the null surface. This generalizes the calculation in \cite{CFP} (see also \cite{Hopfmuller2018,Oliveri:2019gvm}) by weakening the boundary conditions imposed on the field configurations. The expression for the anomalous transformation of the boundary term under diffeomorphisms is derived, and shown to arise from fixing a choice of scaling frame on the null boundary.

\subsection{Geometry of null hypersurfaces} \label{sec:geom}
We start by briefly reviewing the geometric fields on a null hypersurface and their salient properties, following \cite{CFP}. For a detailed review see \cite{Gourgoulhon:2005ng}. Consider a spacetime $(\mathcal{M}, g_{ab})$ and a null hypersurface $\cal N$ in $\cal M$. To begin with, we have the null normal $l_a$ to $\cal N$. An important property of null surfaces is that $l_a$ has no preferred normalization, unlike for spacelike or timelike surfaces. Consequently, we can rescale it according to
\begin{align}
    l_a \rightarrow e^{f}l_a. \label{rescale}
\end{align}
We refer to a choice of $f$ as a \emph{scaling frame}. From $l_a$ we can construct the null generator tangent to $\mathcal{N}$ by raising 
the index, $l^a = g^{ab}l_b$. Associated to the null generator is the inaffinity $k$,\footnote{The inaffinity is often denoted $\kappa$, but we use $k$  to distinguish it from the surface gravity $\kappa$, which is defined on $\NNN$ by the relation
\beq\label{eqn:defnkappa} 
\nabla_a(l^2) \hateq -2\kappa l_a.
\eeq
For Killing horizons, $k = \kappa$, but for general null surfaces, these
two quantities differ; see, e.g.,  \cite{Jacobson:1993pf} for a 
discussion of the difference in the case of conformal Killing horizons.
The definition (\ref{eqn:defnkappa}) of the surface gravity 
is most directly
related to its appearance in the Hawking temperature $T_H = \frac{\kappa}{2\pi}$
\cite{Hawking1974, Gibbons1977c}, which is why we continue to use
$\kappa$ to denote it, and instead use $k$ for the inaffinity.}
defined by
\begin{align}
    l^a \nabla_a l^b \hateq k l^b, \label{inaffinitydef}
\end{align}
where we have introduced the notation $\hateq$ to denote equality at
$\cal N$. The inaffinity will play a central role in this paper. 

We denote by $\Pi^a_i$ the pullback to $\cal N$. Recall that indices $i,j,\ldots$ are intrinsic to $\cal N$. Using the pullback, we can now enumerate the various objects needed for our analysis. The (degenerate) induced metric $q_{ij}$ on $\cal N$ is simply the pullback of $g_{ab}$, 
\begin{align}
    q_{ij} = \Pi^a_i \Pi^b_j g_{ab}. \label{pullbackmetric}
\end{align}
Next, note that $l_b \Pi^a_i \nabla_a l^b \hateq 0$ hence the tensor 
\begin{align}
    \Pi^a_i \nabla_a l^b
\end{align}
is actually intrinsic to $\cal N$. Therefore, we denote it by 
\begin{align}
    S\indices{^i_j}, 
\end{align}
and refer to it as the shape tensor, or Weingarten map \cite{Gourgoulhon:2005ng}. We can extract the inaffinity from the shape tensor through $S\indices{^i_j} l^j = k l^i$. From $S\indices{^i_j}$, we can obtain the extrinsic curvature of $\cal N$, 
\begin{align}
    K_{ij} = q_{jk}S\indices{^k_i}, 
\end{align}
which can be decomposed into its familiar form 
\begin{align}
    K_{ij} = \sigma_{ij} + \frac{1}{d-2}\Theta q_{ij},
\end{align}
where $\sigma_{ij}$ is the shear and $\Theta$ is the expansion. 

Lastly, we can define induced $(d-1)$ and $(d-2)$ volume forms on $\cal N$ as follows. Given a spacetime volume form $\epsilon$, we can define a $(d-1)$ volume form $\tilde \eta$ by
\begin{align}
    \epsilon \hateq - l \wedge \tilde\eta. 
\end{align}
Note that $\tilde\eta$ is fully determined by a choice of $l_a$ up to the addition of terms of the form $l \wedge \sigma$ for some $(d-2)$ form $\sigma$. However, given a choice of $l_a$, the pullback of $ \tilde \eta$ to $\cal N$ is unique. We simply denote this pullback by $\eta$, as we will only be using the pullback henceforth. Given the pullback $\eta$, we can define a $(d-2)$ volume form $\mu$ by
\begin{align}
    \mu = i_l  \eta 
\end{align}
which is uniquely determined by $\eta$. 

We now list the transformation properties of the geometric fields defined above under the rescaling \eqref{rescale}: 
\begin{subequations} \label{rescaleprop}
\begin{equation}
    q_{ij}\rightarrow q_{ij},
\end{equation}
\begin{equation}
    \mu \rightarrow \mu,
\end{equation}
\begin{equation}
    \eta \rightarrow e^{f}\eta, 
\end{equation}
\begin{equation}
    K_{ij}\rightarrow e^{f}K_{ij},
\end{equation}
\begin{equation}
    S\indices{^i_j} \rightarrow e^{f}(S\indices{^i_j} + \partial_j f\, l^i).
\end{equation}
\end{subequations}
We emphasize that this corresponds to a rescaling in a given background geometry. In the next section we will discuss the scale factor $f$ on field space. 

We end this section by introducing an auxiliary null vector $n^a$ on $\cal N$, as it will prove convenient in later computations. We fix the freedom in the relative normalization of $n^a$ by imposing $l_a n^a = -1$. We can use $n^a$ to write the pullback and induced metric as spacetime tensors, 
\begin{subequations}
    \begin{equation}
        \Pi_a^b = \delta_a^b + l_a n^b, \label{spacetimepull}
    \end{equation}
\begin{equation}
    q_{ab} = g_{ab} + 2l_{(a}n_{b)}. \label{spacetimeinduced}
\end{equation}
\end{subequations}
Raising the indices yields a tensor $q^{ab}$ that is tangent to 
$\NNN$ since $q^{ab}l_b = 0$.  It therefore defines a tensor
$q^{ij}$ intrinsic to $\NNN$, which defines a partial inverse
of $q_{jk}$ on the subspace of vectors that annihilate $n_i = \Pi^a_i n_a$.  The mixed index tensor $q\indices{^i_j} = q^{ik}q_{kj}$ is 
then a projector onto this subspace.  

We can also use $n^a$ to define the \Hajicek one-form, 
\begin{align}
    \varpi_a = -q_a^c n^b \nabla_c l_b. \label{Hajicek} 
\end{align}
This pulls back to a one-form $\varpi_i$ on $\NNN$, and under
rescaling (\ref{rescale}), it transforms by
\beq
\varpi_i\rightarrow  \varpi_i + q\indices{^j_i}\partial_j f
\eeq
Using $q^{ij}$ to raise the index of $K_{ij}$, we can give a complete
decomposition of the shape tensor,
\beq
S\indices{^i_j} = l^i(\varpi_j - k n_j) + K\indices{^i_j}.
\eeq
This equation emphasizes the difference between the shape  tensor 
$S\indices{^i_j}$
and the extrinsic curvature $K_{ij}$ on a null hypersurface, 
unlike the case of a
spacelike or timelike hypersurface where the two quantities
have essentially the same content.  An important
point to keep in mind is that the quantities on $\NNN$ that
depend on $n_a$ are $q^{ij}$, $q\indices{^i_j}$, $n_i$,
$K\indices{^i_j}$, and $\varpi_i$, while the quantities
appearing in (\ref{rescaleprop}) are independent of $n_a$.  

\subsection{Boundary conditions} \label{sec:bcs}
We now describe the field configuration space for gravitational theories with a null boundary $\NNN$ in terms of the boundary conditions 
imposed at $\NNN$. An important part of this specification is the choice of a background structure derived from structures defined by
the boundary. A background structure is a set of fields which are constant across the field space. 
Fixing these fields is the source of noncovariance in the gravitational
charge algebra, and ultimately is responsible for the appearance
of central charges. 

To this aim, we start by letting $\cal N$ be a hypersurface
embedded in $\mathcal{M}$, specified by a normal covector field $l_a$. We do not yet impose  that $\NNN$ is a null surface. Consequently, since this specification is independent of the metric, it follows that\footnote{In principle we can allow $l_a$ to rescale under variations according to $\delta l_a \hateq \delta a~ l_a$, but this would unnecessarily introduce an arbitrary non-metric degree of freedom that has no relation to the dynamical degrees of freedom of the theory.}
\begin{align}
    \delta l_a \hateq 0. \label{backgroundstruc}
\end{align}
We take the background structure to solely consist of $l_a$, since all
other quantities relevant for the symplectic
form decomposition are constructed from $l_a$ using the 
metric.\footnote{In particular, we do not impose any constraints on the auxiliary null vector $n^a$, apart from the trivial constraint resulting from fixing the relative normalization $n^a l_a \hateq -1$.}
Now, in order to impose that $\NNN$ is a null surface for all
points in the field space, we must constrain the metric perturbation
$\delta g_{ab}$.
This amounts to the boundary condition 
\begin{align}
    l^a l^b \delta g_{ab} \hateq 0. \label{nullness}
\end{align}
We do not impose any further boundary conditions, so our field configuration space is simply the set of all metrics $g_{ab}$ on a manifold $\cal M$ with boundary $\cal N \subset \partial \cal M$ such that \eqref{backgroundstruc} and \eqref{nullness} are satisfied. This background structure is natural, if not necessary, from the point of view of the gravitational path integral: when we integrate over bulk metrics, we want a null surface as a boundary condition, which must be imposed as a delta function constraint on the dynamical metric, leaving the normal to the surface a non-dynamical variable. 

This is a larger field space than that of \cite{CFP}, where the boundary conditions $\delta k = 0$ and $l^b \delta g_{ab}\hateq 0$ were additionally imposed.  Although both sets of boundary 
conditions lead to the same solution space globally, they differ
from the point of view of the subregion $\sub$, where they represent different choices of boundary degrees of freedom.  
Any additional boundary conditions, beyond the condition (\ref{nullness}) to ensure $\NNN$ is null, eliminate physical degrees of freedom from
the subregion, since these boundary conditions do not correspond
to fixing a degenerate direction of the subregion symplectic form.  Imposing the stronger boundary conditions is equivalent to gauge fixing the global field space using Gaussian null coordinates in the neighborhood of $\NNN$, as was done in various works \cite{Donnay:2016ejv,Chandrasekaran:2019ewn}. As we will see in section \ref{sec:vectorfields}, the diffeomorphisms of interest to us satisfy neither $\delta k =0$ nor $l^b \delta g_{ab}\hateq 0$,  
so we cannot impose these conditions.  In \cite{CFP}, these additional boundary conditions comprised the minimal set necessary for satisfying the Wald-Zoupas stationarity condition $\beom(g_0,\delta g) = 0$ for all $\delta g$, where $g_0$ is a solution in which $\NNN$ is stationary. This stationarity
condition has been argued to be a way of fixing the standard ambiguity
in defining quasiloal charges \cite{Wald2000b, CFP}; 
however, we do not see it as being necessary for the construction to make sense. In its place, we have instead the Dirichlet flux condition \eqref{fluxcond}. Thus, we have imposed the minimal set of boundary conditions needed to specify gravitational kinematics on a manifold with a null boundary.  

We now derive expressions for the variations of $k$ and $\Theta$, which will be needed in the next section when decomposing the symplectic potential. To begin with, we note that\footnote{In \cite{Hopfmuller2018} the $l^a$ component of $\delta l^a$ was made to vanish by 
relaxing the condition $\delta l_a = 0$, instead setting it to $\delta l_a
= -n^b l^c \delta g_{bc} l_a$.  Doing this requires a different fixed 
background structure, which amounts to fixing $n_c$ on the horizon.  
Since they impose no additional constraints on the metric variation, 
the field space in \cite{Hopfmuller2018} is the same as ours, but their analysis
differs in the choice of background structure.  }
\begin{align}
    \delta l^a \hateq (l^b n^c \delta g_{bc})l^a - q^{ab} \delta g_{bc} l^c. \label{varupper}
\end{align}
Using the definition $\Theta = q^{ab}\nabla_a l_b$ of the expansion, and the decomposition \eqref{spacetimeinduced}, we find 
\begin{align}
    \delta \Theta = -\left(\sigma^{ab}+\frac{\Theta}{d-2}q^{ab} \right)\delta g_{ab} - 2 l_c \delta \Gamma^c_{ab}l^a n^b - l_c \delta \Gamma^c_{ab}g^{ab}. 
\end{align}
Separately, using $k = -n^b l^a \nabla_a l_b$, we have 
\begin{align}
\delta k = (k n^b - \varpi^b)l^a \delta g_{ab} + l_c \delta \Gamma^c_{ab}l^a n^b.
\end{align}
In arriving at these expressions we have used that $l_a \delta n^a \hateq -n^a \delta l_a \hateq 0$, which is simply a result of fixing the relative normalization $n^a l_a \hateq -1$ across phase space, combined with $\delta l_a \hateq 0$. In this sense, the expressions for $\delta \Theta$ and $\delta k$ are independent of $\delta n^a$. Thus, combining these two expression, we find 
\begin{align}
    \delta(\Theta + 2k) = 2(k n^b - \varpi^b)l^a \delta g_{ab}-\left(\sigma^{ab}+\frac{\Theta}{d-2}q^{ab} \right)\delta g_{ab} - l_c \delta \Gamma^c_{ab}g^{ab}. \label{bdytermvar}
\end{align}
Lastly, the variation of $\eta$ is given by 
\begin{align}
    \delta \eta = \frac{1}{2}g^{ab} \delta g_{ab}\,\eta \label{volvar}
\end{align}

\subsection{Symplectic potential}
So far we have only discussed the kinematics, which is valid for any theory of gravity. We now take our theory of gravity to be general relativity. By restricting the field space to on-shell configurations, i.e.\ metrics which solve Einstein's equations, we can obtain the associated covariant phase space $\cal P$ as outlined in section \ref{covphase}. The symplectic potential current in general relativity pulled back to $\cal N$ can be written   (momentarily setting $16\pi G = 1$)
\begin{align}
    \boldsymbol{\theta} = \eta \left(\frac{1}{2}l^c\nabla_c
    \left(g^{bc}\delta g_{bc} \right)- l_a g^{bc}\delta \Gamma^a _{bc}\right), \label{eqn:thetapb}
\end{align}
where the bolded tensor $\p\theta$ indicates that it has been pulled back to $\cal N$. We wish to decompose the above expression into boundary, corner, and flux terms, according to the general construction 
described in section \ref{sec:quasicharge}. 

We start by noting that $d\mu = \Theta \eta$.
Using this relation, we have
\begin{align}
    d\left(\frac{1}{2}g^{ab}\delta g_{ab} \,\mu \right) \hateq \frac{1}{2}l^c\nabla_c(g^{ab}\delta g_{ab}) \,\eta + \frac{1}{2}\Theta g^{ab}\delta g_{ab}\, \eta. \label{totalderinter}
\end{align}
The second and first terms in \eqref{eqn:thetapb} appear explicitly in \eqref{bdytermvar} and \eqref{totalderinter} respectively, so we can simply solve for them using these relations. Combining this with \eqref{volvar}, we can write the symplectic potential as
\begin{align}
    \boldsymbol{\theta} = \delta \Big[ (\Theta + 2k)\eta \Big] + d\left[\frac{1}{2}g^{ab}\delta g_{ab} \mu \right] + \eta \left[\sigma^{ab}\delta g_{ab} + 2\varpi^a l^b \delta g_{ab} - \left(k - \frac{\Theta}{d-2}\right)q^{ab}\delta g_{ab} - \Theta g^{bc}\delta g_{bc} \right].
\end{align}
We can shift the $\Theta$ contribution in the boundary term into the corner term by noting that $\delta(\Theta \eta) = d\delta \mu$. Note that this shift 
is an example of an additional ambiguity in the decomposition (\ref{eqn:thdecomp})
of $\p\theta$ 
in separating the corner and boundary terms.  In the present context, this shift
will not affect any central charges since $\Theta\eta$ is covariant, but
in principle this ambiguity can be resolved using the corner improvements 
discussed in appendix \ref{app:cornerimprove}.  

Finally, by making use of \eqref{varupper} we arrive at our desired decomposition of the symplectic potential: 
\begin{align}
    \boldsymbol{\theta} = -\delta \ell + d\beta + \pi^{ij}\delta q_{ij} + \pi_i \delta l^i,  \label{decomp}
\end{align}
where, restoring the factors of $16\pi G$, the various terms in the decomposition are 
\begin{align}
    \ell &= -\frac{k \eta}{8\pi G}, \label{bdy} \\
    \beta &= \frac{1}{16\pi G}(\eta_a \delta l^a + g^{ab}\delta g_{ab}\mu), \label{corner} \\
    \pi^{ij} &= \frac{\eta}{16\pi G}\left[\sigma^{ij} - \left(k + \frac{d-3}{d-2}\Theta\right)q^{ij} \right], 
    \label{eqn:piij}\\
    \pi_i &= -\frac{\eta}{8\pi G}(\varpi_i + \Theta n_i).
    \label{eqn:pii}
\end{align}
This decomposition of the symplectic potential 
on a null boundary is essentially equivalent  to the one 
found in \cite{Parattu_2016}, while it differs slightly
from the expressions in \cite{Hopfmuller2017a, Hopfmuller2018,
CFP} due to differences in choices of boundary conditions.  

The flux terms in \eqref{decomp} are in Dirichlet form, as required by our general prescription. 
The quantity $\pi^{ij}$ defines the conjugate momenta to $\delta q_{ij}$,
the horizontal components of the variation of the induced degenerate
metric on $\NNN$.  The $\frac{d(d-3)}{2}$ components of the shear
make up the momenta associated with gravitons, while the 
scalar $k+\frac{d-3}{d-2} \Theta$ is a scalar momentum
identified in \cite{Hopfmuller2018} as a gravitational pressure.
The other momenta $\pi_i$ are conjugate to  $\delta l^i$.
It can further be decomposed into a vector piece constructed from
the \Hajicek form $\varpi_i$ conjugate to spatial variations of $l^i$,
and a scalar energy density constructed from $\Theta$, 
conjugate to variations
that stretch $l^i$.  Together, $\pi^{ij}$ and $\pi_i$ comprise 
the null analog of the Brown-York stress tensor, which is usually 
defined for timelike hypersurfaces \cite{Brown:1992br}.\footnote{A slightly
different construction in \cite{Aghapour2018, Jafari:2019bpw} found a null
Brown-York stress tensor without the scalar component of $\pi_i$,
but with an additional component conjugate to deformations 
that violate the nullness condition $l^a l^b\delta g_{ab} =0$.
Another approach by \cite{Donnay2019a} obtained a 
null boundary stress tensor as a limit of the 
Brown-York stress tensor on the stretched horizon.  
Their expression differs somewhat from the one presented here. 
}

We now discuss the dependence of the terms in the decomposition on arbitrary choices of background quantities. In writing \eqref{decomp} we introduced a choice of auxiliary null normal $n^a$. Fixing the relative normalization of $n^a$ still leaves the freedom $n^a \rightarrow n^a + V^a + \frac{1}{2}V^2 l^a$, where $V^a$ is any vector such that $n_a V^a = l_a V^a =0$. However, both the boundary term \eqref{bdy} and corner term \eqref{corner} are manifestly independent of $n^a$ hence it follows that the flux term is independent of $n^a$, since $\p\theta$ must be. While the total flux term is independent of $n^a$, $\pi^{ij}$ and $\pi_i$ will in general transform into one another under a change of $n^a$. 

While we have fixed the fluctuation of the scale factor $f$ when defining our phase space, 
we still would like to characterize how various quantities depend
on its background value.  From \eqref{rescaleprop}, we have the following transformation properties of the various terms in the decomposition \eqref{decomp} under a background rescaling: 
\begin{subequations}\label{decomprescale}
\begin{equation}
    \ell \rightarrow \ell - \frac{\eta}{8\pi G}l^i\partial_i f, 
\end{equation}
\begin{equation}
    \pi^{ij} \rightarrow \pi^{ij} - \frac{\eta q^{ij}}{16\pi G} l^k\partial_k f ,     
\end{equation}
\begin{equation}
    \pi_i \rightarrow e^{-f}\left(\pi_i - \frac{\eta}{8\pi G}\partial_i f \right).  
\end{equation}
\end{subequations}

\subsection{Anomalous transformation of boundary term}
\label{sec:anomtransform} 
Having fixed the boundary term, we now derive its noncovariance under diffeomorphisms. We will find that it transforms anomalously, with the anomaly arising directly from fixing a choice of scaling frame \eqref{backgroundstruc}. To see this, we first compute $\lie_{\xi}l_a$ when $\xi^a$ is tangent to $\cal N$, i.e. $\xi^b l_b \hateq 0$. We have 
\begin{align}
    \lie_{\xi}l_a \hateq 2\xi^b \nabla_{[b}l_{a]} + \nabla_a (\xi^b l_b).
\end{align}
Hypersurface orthogonality implies that $\nabla_{[b}l_{a]} \hateq v_{[b}l_{a]}$ for some $v_a$. Moreover, $\nabla_a (\xi^b l_b) \propto l_a$ on $\cal N$. Therefore, 
\begin{align}
    \lie_{\xi}l_a \hateq w_{\xi}l_a. 
\end{align}
Recall that the anomaly operator is defined as $\Delta_{\hat \xi} = L_{\hat \xi} - \lie_{\xi}$. Therefore, since $\delta l_a=0$, we find $\Delta_{\hat\xi}l_a \hateq -w_{\xi} l_a$. 

We also need the noncovariance of the induced volume element. Since $\epsilon$ depends only on the metric, $\Delta_{\hat\xi}\epsilon = 0$. Therefore, using \eqref{volvar}, we just have 
\begin{align}
    \Delta_{\hat\xi}\eta = w_{\xi}\eta.  
\end{align}
Moreover, applying the anomaly operator to $l^b \nabla_b l_a = k l_a$, we find   
\begin{align}
    \Delta_{\hat \xi}k = -w_{\xi}k - l^a \nabla_a w_{\xi}
\end{align}
Putting things together, we have the anomalous transformation of the boundary term: 
\begin{align}
    \Delta_{\hat\xi}(k \eta) = -(l^c\nabla_cw_{\xi})\eta \label{bdyanomaly}
\end{align}
This is one of the main results of this paper. From \eqref{eqn:Kxizeta}, we see that the non-vanishing of the central charge is a consequence of choosing $l_a$ to be the background structure. We discuss the significance of this in section \ref{discussanomaly}. In section
\ref{sec:centralcharges}, we evaluate this anomaly explicitly for the Virasoro generators on a Killing horizon. 

The expression (\ref{eqn:BTbracket}) for the Barnich-Troessaert
bracket that we employ in the next section applies when $\beta$ is covariant, without needing the corner improvements discussed in appendix \ref{app:cornerimprove}. It is easy to see that our choice of corner term (\ref{corner}) does in fact satisfy this. First note that $\Delta_{\hat\xi}\mu = 0$, which handles the second 
term in (\ref{corner}). For the first term, we have 
$ \Delta_{\hat\xi}(\eta_a\delta l^a) = 
(\Delta_{\hat\xi} \eta_a)\delta l^a 
+ \eta_a\delta \Delta_{\hat\xi} l^a
=w_\xi \eta_a\delta l^a - \eta_a\delta(w_\xi l^a) = 0$,
since $\delta w_\xi =0$. 
It follows that the corner term is covariant, $\Delta_{\hat\xi}\beta = 0$, as desired.

As a final note, the fact that the central charge can be expressed 
as a trivial field-dependent cocycle \cite{Barnich:2011mi} according
to (\ref{eqn:trivcocycle}) means that there always exists 
a choice of the flux and boundary terms 
that makes any extensions in the quasilocal
charge algebra vanish.  
Moreover, this choice of flux term would be covariant and rescaling invariant, and was the choice used in \cite{Hopfmuller2018, Adami2020a}. However, consider what would happen if a similar
choice were made for asymptotic symmetries: for example, for $\text{AdS}_3$ asymptotics,
one can choose a boundary term other than the Gibbons-Hawking-York term, in which case the Brown-Henneaux analysis would produce a central charge 
with $c\neq \frac{3R}{2G}$, with $R$ the $\text{AdS}$ radius 
 \cite{Brown1986b}. The flux term in these cases no longer corresponds to Dirichlet boundary conditions. 
In holographic setups, these modified boundary conditions 
lead to CFTs coupled to dynamical metrics \cite{Compere2008}, 
producing complications that are usually avoided in
standard AdS/CFT with Dirichlet boundary conditions.
We therefore draw inspiration from AdS/CFT in imposing that the flux term take Dirichlet form, complementary to the path integral argument in section \ref{sec:quasicharge}.

\subsection{Stretched horizon} \label{sec:stretched}
\begin{figure}
\centering
\includegraphics[width=0.4\textwidth]{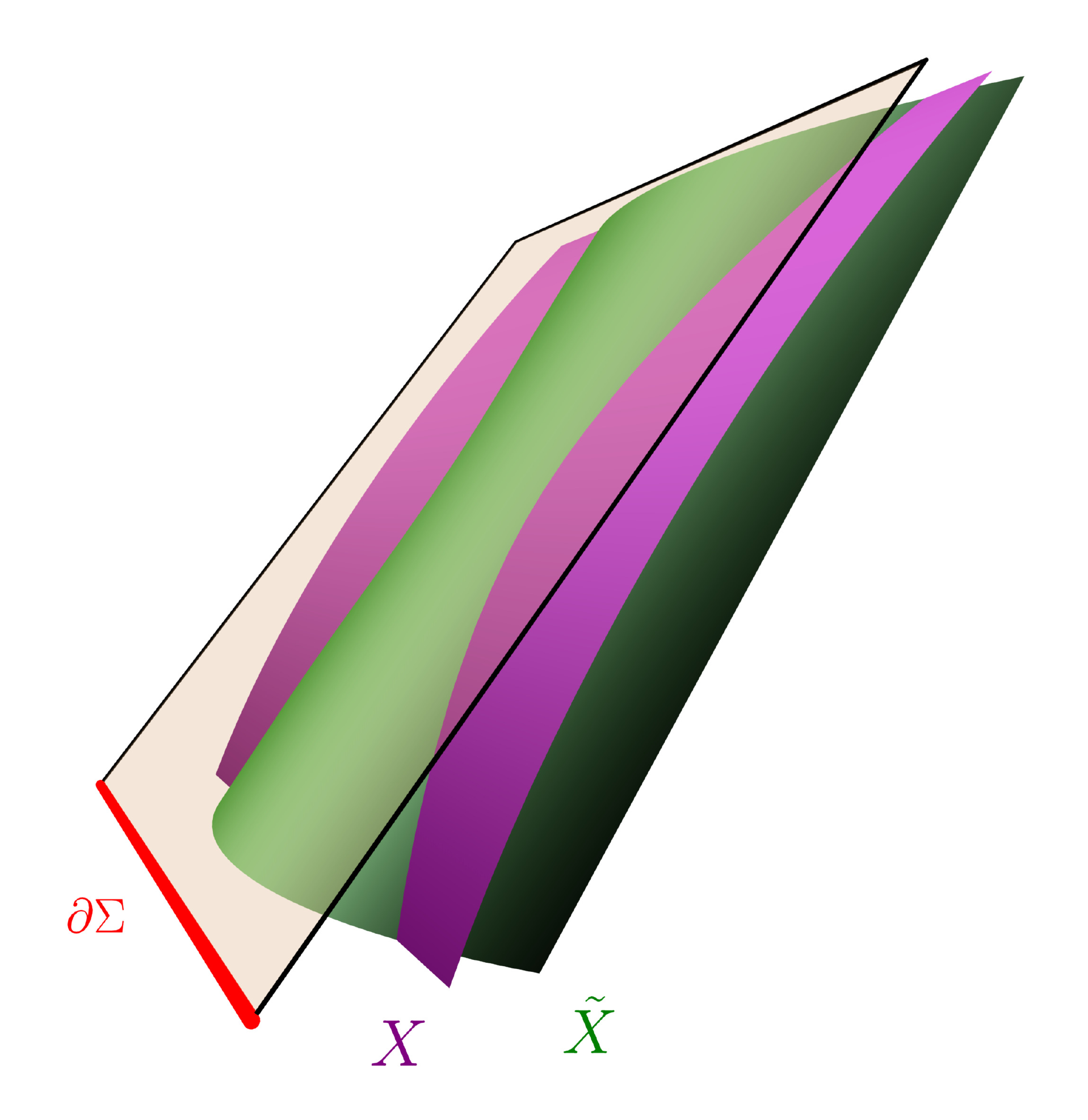}
\caption{Two different choices of stretched horizons are shown,
as the level sets of the functions $X$ and $\tilde X$,
which lead to different scaling frames for $l_a$ on the null 
surface. \label{fig:stretch}}
\end{figure}

We mentioned in section \ref{sec:geom} 
that fixing $l_a$ corresponds to 
a type of frame choice.  Here, we will relate this choice 
to the arbitrariness in choosing a sequence of stretched horizons
that approach the null surface.  A stretched horizon for a null surface
plays a similar role to an asymptotic cutoff surface when discussing 
asymptotic infinity.  These are especially relevant in AdS/CFT,
where different choices of the radial cutoff correspond to different 
conformal frames in the dual theory.  This then strengthens the 
relation between the scaling frame for $l_a$ and the choice of 
conformal frame for the degrees of freedom associated with the 
quasilocal charges.  

To see the relation, we let $X$ denote a function whose level
sets define the sequence of stretched horizons approaching $\NNN$
at $X=0$.  We let $l_a$ be the (unnormalized) normal form to the 
$X$ foliation,
\beq
l_a = \nabla_a X,
\eeq
which is spacelike for $X>0$ and null at $X=0$.  Any reparameterization
of the form $X\rightarrow \tilde{X}(X)$ defines the same foliation, 
and its effect on the normal is 
simply to rescale $l_a$ by $\tilde{F}'(X)$.  Hence $l_a$ at $\NNN$ only rescales
by a constant $\tilde{X}'(0)$.  We therefore see that the scaling frame of 
$l_a$ is determined by the choice of stretched horizon foliation, up 
to overall constant rescalings.  

A different foliation of stretched horizons can be obtained by 
reparameterizing by an arbitrary function of the 
coordinates $X\rightarrow \tilde{X}(X, x^i)$, subject to the constraint
$\tilde{X}(0,x^i) = 0$, so that the foliation still approaches $\NNN$ (see figure \ref{fig:stretch}).  The 
null normal is now rescaled by the position dependent function 
$\partial_{X} \tilde{X}(0,x^i)$, corresponding to a change of scaling frame.  

\section{Virasoro symmetry }\label{killinghorizon}
As an application of the null boundary covariant phase space we have just constructed, 
we now specialize to the case of bifurcate, axisymmetric Killing horizons.  
These have been the subject of many previous analyses, in which  
quasilocal charge algebras have been used to derive  expressions for the 
entropy of the Killing horizon 
\cite{ Carlip_1999,Haco2018,Strominger_1998, Carlip_1999a,
Guica2009, Carlip2011b}.  The standard procedure is to find 
a set of vector fields in the near-horizon region whose Lie brackets 
yield one or two copies of the Witt algebra.  Upon computing the
quasilocal charge algebra, one generally finds a central extension.  
The resulting Virasoro algebra is the symmetry algebra
of a 2D CFT, suggesting that the quantization of the near horizon 
charge algebra should have a CFT description.  
The asymptotic density of states in such a theory is controlled 
by the Cardy formula, and by applying it in conjunction with the 
central charge computed from the quasilocal charge algebra, one arrives at the Bekenstein-Hawking entropy. 

This procedure for arriving at the horizon entropy has been applied in 
a variety of different situations, often differing in the precise 
details of which symmetry algebra is used and what boundary conditions
are imposed 
\cite{Bagchi2013, Afshar2016, Carlip2018a, Carlip2020, Aggarwal2020}.  Here, by means of example, we provide evidence for the claim that the central charge occurring in these setups is always 
computed by the general formula (\ref{eqn:Kxizeta}) 
in terms of the noncovariance
of the boundary Lagrangian for the null surface.  The example we will
analyze is the set of symmetry generators found for axisymmetric
Killing horizons in \cite{Chen:2020nyh}, which generalize the 
near horizon conformal symmetries of the Kerr black hole proposed by
Haco, Hawking, Perry, and Strominger (HHPS) \cite{Haco2018}.  We show that the 
null surface Wald-Zoupas construction described above
produces a formula 
for the central charge which, via the Cardy formula,
leads to an entropy that is twice the Bekenstein-Hawking
entropy of the horizon.  
We argue that this factor of 2 could arise if the central
charge was sensitive to both sets of edge modes, one
on either side of the bifurcation surface, coupled 
together by the Dirichlet flux matching condition.  To 
make a contradistinction, we compare to the 
case where boundary conditions are found to make the 
quasilocal charges integrable, and show that a different 
central charge results, and no factor of 2 appears.  
This thereby gives a derivation of the appropriate
``counterterms'' (i.e.\ fluxes)  that had previously been conjectured
to be necessary for the construction in \cite{Haco2018, Chen:2020nyh}.

\subsection{Near-horizon expansion} \label{sec:near-horizon}
We begin by reviewing the expansion of the metric near a bifurcate 
Killing horizon, following a construction of Carlip 
\cite{Carlip_1999a, Chen:2020nyh}.  Let $l^a$ be the horizon-generating Killing vector, which is timelike
in the exterior region, and becomes the null normal on the bifurcate
Killing horizon $\HHH$.  A canonical choice of radial vector can be 
made using the gradient of the norm of $l^a$, 
\beq \label{eqn:rhoa}
\rho^a = -\frac{1}{2\kappa} \nabla^a\lsq, 
\eeq
where $\kappa$ is the surface gravity, which is constant on account 
of the zeroth law of black hole mechanics \cite{Bardeen1973a}.  
The normalization
of $\rho^a$ is chosen so that it coincides with $l^a$ on $\HHH$, and as 
a consequence of Killing's equation, one finds that $l\cdot \rho = 0$ and 
$[l,\rho] = 0$ everywhere.  If in addition the horizon is axisymmetric,
meaning it possesses a rotational Killing vector $\psi^a$ that commutes
with $l^a$, it follows that $\psi \cdot \rho = 0$ and $[\psi, \rho] =0$.  
This allows us to choose coordinates $(t, r_*, \phi)$ such that 
$(l^a, \rho^a, \psi^a)$ are the corresponding coordinate basis vectors,
and in this coordinate system, $g_{t r_*} = g_{\phi r_*} = 0$.  
 The 
radial coordinate $r_*$ is analogous to the tortoise coordinate in 
the Schwarzschild solution, with the horizon positioned at $r_*\rightarrow
-\infty$.  The remaining  coordinates will be denoted $\theta^A$. 

One can demonstrate that the norm of the radial vector 
near the horizon satisfies \cite{Carlip_1999a}
\beq \label{eqn:l2rho2}
\rho \cdot \rho = -\lsq + \mathcal{O}\left[\lsq^2\right],
\eeq
and hence as a function of $r_*$, the Killing vector norm satisfies
the differential equation
\beq
\partial_{r_*} \lsq = \rho^a\nabla_a\lsq 
= 2\kappa \lsq +\mathcal{O}\left[\lsq^2\right]
\eeq
whose solution is 
\beq
\lsq =  -e^{2\kappa r_*} + 
\mathcal{O}\left[ e^{4\kappa r_*}\right],
\eeq
where the integration constant has been absorbed by the shift freedom in the 
definition of the tortoise coordinate, $r_* \rightarrow r_* + f(\theta^A)$. 
This behavior suggests a reparameterization of the radial coordinate,
\beq\label{eqn:x}
x = \frac1\kappa e^{\kappa r_*},\quad \implies
\quad \partial_x^a = \frac{1}{\kappa x}\rho^a
\eeq
in terms of which the Killing vector norm has the expansion
\beq
\lsq = -\kappa^2 x^2 + \mathcal{O}\left[x^4\right].
\eeq
This also implies that $\partial_x^a$ is unit normalized to leading 
order in the near-horizon expansion, which means $x$ coincides
with the radial geodesic distance to the bifurcation surface 
at this order.  This fully determines the $x$ coordinate,
and in terms of it, the near-horizon metric exhibits a Rindler-like 
expansion,
\beq
ds^2 = -\kappa^2 x^2 dt^2 + dx^2 + \psi^2 d\phi^2 + 
q_{AB} d\theta^A d\theta^B 
-2x^2\kappa dt\left(N_\phi d\phi + N_A d\theta^A\right) + \ldots
\eeq
where the $\ldots$ denotes higher order terms which do not play a role in the remainder 
of the analysis of the near horizon symmetries.  Here,
we have used the shift freedom $\phi\rightarrow \phi + G(\theta^A)$ to 
eliminate any $d\phi d\theta^A$ terms that generically
appear. 

The Rindler coordinates degenerate on the future and past 
horizons, so it is useful to define Kruskal coordinates which are regular
on the horizon,
\begin{subequations}\label{eqn:Kruskal}
\begin{align}
U &= -x e^{-\kappa t} \\
V &= x e^{\kappa t},
\end{align}
\end{subequations}
in terms of which the metric becomes
\beq
ds^2 = -dU dV + \psi^2 d\phi^2 + q_{AB}d\theta^A d\theta^B 
+(UdV-VdU)(N_\phi d\phi + N_A d\theta^A) +\ldots
\eeq
The Killing vector and radial vector have simple expressions in
terms of Kruskal coordinates,
\begin{align}
l^a &= \kappa(V\partial_V^a - U\partial_U^a) \label{eqn:la} \\
\rho^a &= \kappa(V\partial_V^a + U\partial_U^a),
\end{align}
which demonstrates that near the bifurcation surface at $U=V=0$,
$l^a$ acts like a boost while $\rho^a$ acts like a dilatation.  

The future horizon $\mathcal{H}^+$
in Kruskal coordinates is located at $U=0$, and on the 
horizon the generator is  $l^a = \kappa V\partial_V^a$.  The natural
choice of auxiliary null covector there is then 
$n_a = -\frac1{\kappa V}\nabla_a V + \frac12
\left|\frac{dV}{\kappa V}\right|^2 l_a$, where the term proportional
to $l_a$ just ensures that $n_a$ is null on all of $\mathcal{H}^+$.  
The spacetime volume form is given by
\beq
\ep = \frac12 dU\wedge dV \wedge \mu = -l\wedge \eta,
\eeq
where the induced volume form on the horizon is 
\beq
\eta = \frac{1}{\kappa V} dV\wedge \mu.
\eeq

The past horizon $\mathcal{H}^-$
is at $V=0$, where the generator is $l^a = -\kappa U\partial_U^a$ and 
the auxiliary null covector is $n_a = \frac{1}{\kappa U}\nabla_a U+
\frac12\left|\frac{dU}{\kappa U}\right|^2 l_a$.  The conventions we 
use to define the volume forms are slightly different than on 
the future horizon.  We choose the volume form on the past horizon 
to be 
\beq
\eta = -\frac{1}{\kappa U} dU \wedge \mu,
\eeq
to maintain the relationship $\mu = i_l\,\eta$.  This means that the 
spacetime volume is related to $\eta$ on the past horizon by
\beq\label{eqn:ep-}
\ep = l\wedge \eta,
\eeq
and these conventions ensure that $\mu$ limits to the same volume 
form on the bifurcation surface when approached on  
$\mathcal{H}^+$ or on $\mathcal{H^-}$.  Because of (\ref{eqn:ep-}), 
the decomposition of $\p\theta$ picks up an overall minus sign 
relative to the expression (\ref{eqn:thetapb}).  This means that on 
$\mathcal{H}^-$, the boundary term has a relative minus sign
compared to (\ref{bdy}) 
\beq
\ell = \frac{k\eta}{8\pi G}\qquad (\text{on $\mathcal{H}^-$} ).
\eeq

\subsection{Expression for the noncovariance}
\label{sec:noncovariance}
The results of section \ref{sec:BTbrack} show that any extension of 
the quasilocal symmetry algebra is determined by the noncovariance 
of the boundary term, $\Delta_{\hat\xi} \ell$.  The noncovariance of this
quantity and the various other structures defined on a generic null
surface were determined  in section \ref{sec:anomtransform} in terms of the 
scalar $w_\xi$ which shows up in the noncovariance of the normal 
form to the horizon, $l_a$.  To apply these formulas in computations
of the algebra extensions, we therefore need an expression
for $w_\xi$ on a Killing horizon.  

This can be derived on $\mathcal{H}^+$ 
by first noting that if $\xi^a$ is tangent to 
the null surface $\NNN = \mathcal{H}^+$, 
the value of $\lie_\xi l_a$ does not 
depend on how $l_a$ is chosen away from $\NNN$.  Since $l_a$ and $\rho_a$ 
coincide on $\NNN$, we can compute 
$w_\xi l_a = \lie_\xi l_a \hateq \lie_\xi \rho_a = \nabla_a(\xi \cdot \rho)$,
since $(d\rho)_{ab} = 0$ due to its definition as a gradient in equation (\ref{eqn:rhoa}).  
To continue the calculation, we express $\xi^a$ 
in terms of the basis $(l^a, \rho^a, \psi^a, \partial_A^a)$ 
as $\xi^a = \xi^\rho \rho^a + V^a$, where $V^a$ is some 
combination of $l^a$, $\psi^a$, and $\partial_A^a$.  Since 
$l\cdot \rho = \psi \cdot \rho = 0$ everywhere, and $\partial_A \cdot \rho
= \mathcal{O}[x^3]$, when evaluated on the horizon, only the $\xi^\rho$
component survives in the gradient.  Hence we find, using 
(\ref{eqn:l2rho2}),
\beq
\nabla_a(\xi\cdot \rho) \hateq \xi^\rho \nabla_a(\rho\cdot \rho)
\hateq -\xi^\rho \nabla_a \lsq \hateq 2\kappa \xi^\rho\, l_a.
\eeq
This leads to the simple expression,
\beq \label{eqn:wxi+}
w_\xi = 2\kappa \xi^\rho \qquad \text{(on $\mathcal{H}^+$)},
\eeq
so we see that the noncovariance comes entirely from the dilatation
component of $\xi^a$, i.e.\ the component parallel to $\rho^a$.  
Note that although $w_\xi$ does not depend on how $l_a$ is extended 
off of $\NNN$, it does depend on the extension of $\xi^a$ in the vicinity
of $\NNN$.  To demonstrate this point, we note that because $l^a$ and 
$\rho^a$ coincide on $\NNN$, one cannot separate $\xi^a$ into its 
$l^a$ and $\rho^a$ components using  its value on $\NNN$ alone. Only after 
looking at its behavior as you move away from $\NNN$ can its $l^a$ and 
$\rho^a$ components be distinguished, and then only the $\rho^a$ 
component contributes to the noncovariance. 

The analysis on the past horizon $\mathcal{H}^-$ is similar 
and leads to
\beq \label{eqn:wxi-}
w_\xi = 2\kappa \xi^\rho \qquad \text{(on $\mathcal{H}^-$)}.
\eeq

\subsection{Virasoro vector fields} \label{sec:vectorfields}

Having introduced the near-horizon expansion of the metric,
we now turn to the choice of vector fields generating the near-horizon
symmetries.   Motivated by the hidden conformal symmetry of scattering
amplitudes in Kerr \cite{Castro2010}, HHPS proposed
a set of vector fields for Kerr black holes whose algebra consisted of 
two commuting copies of the Witt algebra.  This algebra was identified
by foliating the near-horizon region by approximately $\text{AdS}_3$ slices,
and writing down the corresponding asymptotic symmetry generators.
The construction of these symmetry generators was extended
to Schwarzschild black holes in \cite{Averin2020}, which
also proposed a two-parameter generalization 
in the choice of vector fields, with
the two parameters coinciding with notions of left and right temperatures.
The construction was further extended to arbitrary axisymmetric 
Killing horizons in \cite{Chen:2020nyh}, which similarly identified an
algebra $\text{Diff}(S^1)_\alpha \times \text{Diff}(S^1)_{\bar\alpha}$,
consisting of two commuting copies of the Witt algebra, and labeled by
two parameters $(\alpha,\bar\alpha)$ which coincide with choices of 
temperatures.  In this section, we will analyze this latter algebra
for general choices of $(\alpha,\bar\alpha)$, and show 
in section \ref{sec:centralcharges} that the quasilocal charge algebra leads to an expression
for the central charges.  

 One way to describe the symmetry algebra is to present it in terms of a 
geometric structure that it preserves.  To this end, we define the 
following ``conformal coordinates'' depending on the two
parameters $(\alpha,\bar\alpha)$ \cite{Chen:2020nyh}:
\begin{subequations} \label{eqn:confcoords}
\begin{align}
W^+ &= V e^{\alpha\phi} \label{eqn:Wp}\\
W^- &= -U e^{\bar\alpha\phi} \\
y &= e^{\frac{\alpha+\bar\alpha}{2} \phi}.
\end{align}
\end{subequations}
The $2\pi$ periodicity of  $\phi$  
requires that these coordinates be identified according to
$(W^+, W^-, y)\sim (e^{2\pi \alpha} W^+, e^{2\pi\bar\alpha} W^-, e^{\pi(\alpha+\bar\alpha)} y)$.  We then form the following tensor
\beq \label{eqn:Cab}
C_{ab} = -\frac1{y^2}\nabla_a W^+\, \nabla_b W^- =
\Big(\nabla_a V + \alpha V \nabla_a\phi\Big)
\Big(\nabla_b U + \bar\alpha U \nabla_b \phi\Big)
\eeq
where the second equality demonstrates that $C_{ab}$ is well-defined
in light of the periodicity of the conformal coordinates.  The 
near-horizon symmetries are defined to simply be the transformations
that preserve $C_{ab}$.  A trivial set of such transformations are simply
those parallel to the transverse directions, $V^A\partial_A$.  They 
preserve the bifurcation surface of the horizon, and hence do not 
require the Wald-Zoupas prescription, nor do they lead to algebra
extensions when represented in terms of quasilocal charges.  We therefore
focus on the nontrivial transformations that act in the $(t, r_*,\phi)$
plane.  

Using the first expression for $C_{ab}$ in (\ref{eqn:Cab}), it is 
straightforward to see that the vector fields that satisfy 
$\lie_\xi C_{ab}=0$ are of the form
\begin{align}
\xi^a_n &= F_n(W^+) \partial_+^a + \frac12 F_n'(W^+) y\partial_y^a 
\label{eqn:xinconf}
\\
\bar\xi^a_n&= \bar F_n(W^-) \partial_-^a+\frac12 \bar F'_n(W^-) y \partial_y^a. 
\label{eqn:xinbarconf}
\end{align}
In order to be single-valued, the functions $F_n, \bar F_n$ must 
satisfy $F_n(W^+ e^{2\pi\alpha}) = F_n(W^+) e^{2\pi\alpha}$, 
$\bar F_n(W^- e^{2\pi\bar \alpha}) = \bar F_n(W^-) e^{2\pi\bar \alpha}$,
and hence they can be expanded in modes,
\begin{align}
F_n &= \alpha W^+ \left(W^+\right)^{\frac{in}{\alpha}} \\
\bar F_n &= -\bar\alpha W^- \left(W^-\right)^{-\frac{in}{\bar \alpha} }.
\end{align}
We can then compute the Lie brackets of these vector fields, and 
find that their algebra is given by two commuting copies of 
the Witt algebra,
\begin{align}
[\xi_m, \xi_n] &= i(n-m) \xi_{m+n} \\
[\bar \xi_m,\bar \xi_n] &= i(n-m)\bar \xi_{m+n} \\
[\xi_m,\bar \xi_n] &= 0
\end{align}

Although preservation of the tensor $C_{ab}$ uniquely specifies 
the near-horizon symmetry generators, there is still a question
as to why this is a useful criterion to impose.  While we do not 
have a completely satisfactory answer, we can point out some interesting
features of $C_{ab}$ that may inform future investigations into 
its significance.  First we note that the vector fields also preserve 
the following contravariant tensor,
\beq
D^{ab} = -y^2 \partial_+^a \partial_-^b = 
\partial_V^a \partial_U^b =
\frac{1}{2\kappa^2 x^2}(l^a +\rho^a)(l^b-\rho^b),
\eeq
for any choice of $(\alpha,\bar\alpha)$. From this, one can also 
construct the projectors 
\begin{align}
(P_+)\indices{^b_a} &= C_{ac}D^{bc} = \nabla_a W^+ \partial^b_+ =\left(\frac{\nabla_a V}{\kappa V}
+ \frac{\alpha}{\kappa} \nabla_a \phi\right) \kappa V \partial_V^b\\
(P_-)\indices{^b_a} &= C_{ca}D^{cb} = \nabla_a W^- \partial_-^b
=
\left(\frac{\nabla_a U}{\kappa U}+\frac{\bar\alpha}{\kappa} 
\nabla_a \phi\right) \kappa U \partial_U^b
\end{align}
which are also preserved.  On $\mathcal{H^+}$, the upper index
of $(P_+)\indices{^a_b}$ is parallel to the horizon generator, and 
so by pulling back the lower index to $\mathcal{H^+}$, one arrives
at a vertical projector for vectors on $\mathcal{H^+}$ onto $l^a$.  Such
a projector is an example of an Ehresmann connection for the horizon, 
viewed as a fiber bundle with fibers consisting of the null flow lines 
of $l^a$.  It is, in fact, a flat connection, with horizontal directions given
by the surfaces of constant $W^+$.  However, this connection produces
a nontrivial holonomy upon completing a $2\pi$ rotation in $\phi$, 
which results in $V\rightarrow Ve^{-2\pi\alpha}$ (see \cite{Chen:2020nyh}
for a depiction of this spiraling behavior of the conformal coordinates).
$(P_-)\indices{^a_b}$ similarly defines a flat Ehresmann connection on
the past horizon, with $2\pi$ holonomy $U\rightarrow Ue^{-2\pi\bar\alpha}$.

The relevance of such Ehresmann connections in the study of Carroll
geometries on null surfaces \cite{Duval2014}
 was recently
emphasized in \cite{Ciambelli2019b}, so investigating the relationship
between Carroll geometries and the near-horizon Virasoro symmetries
may lead to a deeper understanding as to their fundamental origin.  
Note, however, it is important that the generators are defined to preserve
$C_{ab}$ in a neighborhood of the bifurcation surface; it is not enough to 
simply find vector fields that preserve $P_+$ and $P_-$ on each of the 
respective horizons.  This is because the behavior of $\xi^a_n$ off of the 
horizon determines the noncovariances, which in turn determine 
extensions of the quasilocal charge algebra.  Since $C_{ab}$ contains
the information about both projectors, the geometric interpretation of 
the symmetry generators seems to involve not only the Ehresmann
connections on each individual horizon, but also how they relate 
to each other in forming a bifurcate horizon.  

As discussed in section \ref{sec:noncovariance}, the noncovariances 
depend on the $\rho^a$ component
of the symmetry generators.  This can be computed by transforming
the vector fields (\ref{eqn:xinconf}) and (\ref{eqn:xinbarconf}) back to 
the $(t,r_*,\phi)$ coordinate system, in which they are expressed
in terms of $l^a$, $\rho^a$, and $\psi^a$.  Using (\ref{eqn:x}),
(\ref{eqn:Kruskal}), and (\ref{eqn:confcoords}), this leads to 
\begin{align}
\xi_n^a &= \frac{\left(W^+\right)^{\frac{in}{\alpha}}}{\alpha+\bar\alpha}
\left[\frac{\alpha\bar\alpha}{\kappa} l^a + \alpha \psi^a 
+in\left( \frac{\bar\alpha -\alpha}{2\kappa} l^a +\psi^a\right)\right]
-\frac{in}{2\kappa}\left(W^+\right)^{\frac{in}{\alpha}} \rho^a 
\label{eqn:xin}
\\
\bar\xi_n^a 
&= \frac{\left(W^-\right)^{-\frac{in}{\bar\alpha}}}{\alpha+\bar\alpha}
\left[\frac{\alpha\bar\alpha}{\kappa} l^a - \bar\alpha \psi^a 
+in\left( \frac{\bar\alpha -\alpha}{2\kappa} l^a +\psi^a\right)\right]
-\frac{in}{2\kappa}\left(W^-\right)^{-\frac{in}{\bar\alpha}} \rho^a,
\label{eqn:barxin}
\end{align}
Note that the prefactor  $\left(W^+\right)^{\frac{in}{\alpha}} = V^{\frac{in}{\alpha}} e^{in\phi}$ in $\xi_n^a$ has an oscillating 
singularity as the past horizon at $V\rightarrow 0$ is approached.  
This means that the $\xi_n^a$ vector fields have no well-defined limit
to the past horizon, and so their quasilocal charges will be constructed
on the future horizon.  Similarly, the prefactor
$\left(W^-\right)^{-\frac{in}{\bar\alpha}} = (-U)^{-\frac{in}{\bar\alpha}}
e^{-in\phi}$ in $\bar\xi_n^a$ has no limit to the future horizon
$U\rightarrow 0$, and so the corresponding
quasilocal charges will be evaluated on 
$\mathcal{H}^-$.  With this in mind, we can read off the 
expression for the noncovariances associated with these vector fields 
using (\ref{eqn:wxi+}) and (\ref{eqn:wxi-}), which gives
\begin{align}
w_{\xi_n}&= -in \left(W^+\right)^{\frac{in}{\alpha}} 
\qquad \text{(on $\mathcal{H^+}$)} \label{eqn:wxW+}\\
w_{\bar\xi_n} &= -in \left(W^-\right)^{-\frac{in}{\bar\alpha}} 
\qquad \text{(on $\mathcal{H^-}$)}.
\end{align}

We now demonstrate that these vector fields do not preserve
the boundary conditions $\delta k = 0$, $\delta l^a \hateq 0$, or 
$n_a\delta l^a \hateq 0$ that have been employed in previous
works \cite{CFP,Hopfmuller2018,Donnay:2016ejv, Chandrasekaran:2019ewn}.  On $\mathcal{H}^+$,  
\begin{align}
I_{\hat\xi_n} \delta k &= -n(n-i\alpha) \frac{\kappa}{\alpha} 
\left(W^+\right)^{\frac{in}{\alpha}},
\label{eqn:delxik}\\
I_{\hat\xi_n}\delta l^a &= \frac{n(n-i\alpha)}{\alpha+\bar\alpha}
\left(W^+\right)^{\frac{in}{\alpha}} \left[-l^a +\frac\kappa\alpha
\psi^a\right],
\label{eqn:delxil}
\end{align}
which clearly violates all three conditions pointwise.  
These conditions are also violated pointwise by 
the $\bar\xi_n^a$ generators on $\mathcal{H}^-$,
\begin{align}
I_{\hat{\bar\xi}_n} \delta k &=
-n(n+i\bar\alpha)\frac{\kappa}{\bar\alpha}
\left(W^-\right)^{-\frac{in}{\bar\alpha}} 
\label{eqn:delbxik}\\
I_{\hat{\bar\xi}_n} \delta l^a &=
\frac{n(n+i\bar\alpha)}{\alpha+\bar\alpha} 
\left(W^-\right)^{-\frac{in}{\bar\alpha}} 
\left[l^a +\frac{\kappa}{\bar\alpha} \psi^a\right].
\label{eqn:delbxil}
\end{align} 
This therefore necessitates the use of the weaker boundary conditions
described in section \ref{sec:bcs}.

\subsection{Central charges} \label{sec:centralcharges}
With all this in place, we can proceed to the calculation
of the central extension of the quasilocal charge algebra.
We denote 
the  quasilocal charges for $\xi_n^a$ by $L_n$, and the 
charges for $\bar\xi_n^a$ by $\bar L_n$.  Their values are given 
 by the general expression (\ref{eqn:Hxi}), evaluated on 
 $\mathcal{H}^+$ for the $L_n$ generators and on $\mathcal{H}^-$
 for the $\bar L_n$ generators.  Note that because 
the background is rotationally symmetric, all of the 
charges $L_n$, $\bar L_n$ except for $L_0$, $\bar L_0$ vanish,
since the generators (\ref{eqn:xin}), (\ref{eqn:barxin})
come with angular dependence  $e^{in\phi}$, which integrates 
to zero on $\partial\Sigma$.  
Of course,
their variations, which enter the calculation 
of the brackets, need not vanish.  
Since the vector fields $\xi_0^a$ and $\bar \xi_0^a$ are linear combinations
of the horizon-generating and rotational Killing vectors, $l^a$ and $\psi^a$, 
the $L_0$, $\bar L_0$ charges 
will be linear combinations of the  Noether
charges for the Killing vectors, 
namely, the horizon area $A$ and angular momentum $J_H$. 
The zero mode generators evaluate to
\begin{align}
L_0 &= \frac{\alpha}{\alpha+\bar\alpha} J_H \\
\bar L_0 &= -\frac{\bar\alpha}{\alpha+\bar\alpha} J_H,
\end{align}
where the horizon angular momentum $J_H$ is given by 
the Noether charge for the rotational Killing vector $\psi^a$,
\beq
J_H = \int_{\partial\Sigma} Q_\psi 
= \frac{1}{4G}\int d\theta^A\sqrt{q}|\psi| N_\phi(\theta^A).
\eeq
The area contribution has dropped from these expressions 
because the quasilocal charge $H_l$ 
for $l^a$, which normally is proportional to the area, vanishes
upon including the Dirichlet 
boundary term $i_l \ell$ from (\ref{eqn:Hxi}).  This is somewhat
unintuitive because $l^a$ vanishes as the bifurcation surface
is approached; however, the contraction with $\ell$ has a nonzero
value in the limit.  The vanishing of this boost Noether
charge was similarly observed in the analysis of a phase space
bounded by a timelike hypersurface with Dirichlet boundary 
conditions \cite{Harlow2019, Freidel2020}.

The discussion of 
section \ref{sec:BTbrack}  showed that the Barnich-Troessaert 
bracket of the charges must reproduce the algebra of the 
vector fields, up to abelian extensions.  
Hence, for the $\xi_n^a$ vector fields, 
the bracket of the charges can be written
\beq
\{L_m, L_n\} = -i\Big[ (n-m) L_{m+n} + K_{m,n}\Big],
\eeq
where $K_{m,n}$ is determined by the explicit formula
(\ref{eqn:Kxizeta}),
\beq
K_{m,n} = -i\int_{\partial\Sigma} \Big(i_{\xi_m}\Delta_{\hat\xi_n}
\ell - i_{\xi_n}\Delta_{\hat\xi_m} \ell\Big).
\eeq

To evaluate this, we first note that the  expression 
(\ref{bdyanomaly}) for the 
noncovariance of $k\eta$ and the 
expression (\ref{eqn:wxW+}) for $w_{\xi_n}$ 
gives 
\beq
\Delta_{\hat\xi_n}\ell  = \frac{\eta}{8\pi G} l^a\nabla_a 
w_{\xi_n} = \frac{\eta}{8\pi G} \frac{n^2\kappa }{\alpha} 
\left(W^+\right)^{\frac{in}{\alpha}}.
\eeq
For the quantity $i_{\xi_m}\eta$, note that the $\psi^a$ component
will not contribute to this expression when evaluated on a 
surface of constant $V$.  Recalling that $\rho^a = l^a$ on 
$\mathcal{H^+}$, we have 
\beq
i_{\xi_m} \eta = \frac{\left(W^+\right)^{\frac{in}{\alpha}}}{\alpha+\bar\alpha} 
\left(\frac{\alpha\bar\alpha}{\kappa} - im\frac{\alpha}{\kappa}\right)
i_l\eta = 
\frac{\left(W^+\right)^{\frac{in}{\alpha}} }{\alpha+\bar\alpha}
\frac{\alpha}{\kappa} 
\left(\bar\alpha - im\right)
\mu.
\eeq
Then we find that 
\beq
i_{\xi_m} \left(\Delta_{\hat\xi_{-m}}\ell\right) = -im^2 \frac{(m +i\bar\alpha)}{(\alpha+\bar\alpha) } \frac{\mu}{8\pi G},
\eeq
and subtracting the term with $m \leftrightarrow -m$ and 
integrating over the surface gives a result proportional to the 
horizon area $A$,
\beq
K_{m,-m} =  \frac{A}{4\pi G (\alpha+\bar\alpha)} m^3.  
\eeq
Any other extension term $K_{m,n}$ with $m\neq -n$ vanishes, again
due to rotational invariance and the overall $e^{-i(m-n)\phi}$ 
dependence of the integrand.  We verify in appendix \ref{app:checkingcentral}  that 
the variations of the quantities $K_{m,n}$ with $m\neq -n$ 
are consistent with 
having identically zero quasilocal charges associated with them,
which means that the only nontrivial extension terms are $K_{m,-m}$.
Hence, the extension is in fact central, and the algebra 
obtained is the Virasoro algebra,
\beq
\{L_m,L_n\} = -i\Big[(n-m) L_{m+n} + 
\frac{c}{12}m^3 \delta_{m,-n} \Big]
\eeq
with central charge
\beq\label{eqn:c}
c = \frac{3A}{\pi G(\alpha+\bar\alpha)}.
\eeq

The analysis for the $\bar\xi_n^a$ generators is similar.  The 
calculations need to be done on the past horizon due to the 
singularity in $\bar\xi_n^a$ on the future horizon.  As explained
in section \ref{sec:near-horizon}, this flips the sign of the boundary term $\ell$ 
in the decomposition of the symplectic form.  This then gives
\begin{align}
\Delta_{\hat{\bar{\xi}}_n}\ell 
&= 
-\frac{\eta}{8\pi G} l^a\nabla_a
w_{\bar\xi_n} = -\frac{\eta}{8\pi G} \frac{n^2\kappa}{\bar\alpha} 
\left(W^-\right)^{-\frac{in}{\bar\alpha}} \\
i_{\bar\xi_n} \eta 
&=
\frac{\left(W^-\right)^{-\frac{in}{\bar\alpha}} }
{\alpha+\bar\alpha} \frac{\bar\alpha}{\kappa}(\alpha+in) \mu\\
i_{\bar\xi_m}\left(\Delta_{\hat{\bar{\xi}}_{-m}}\ell \right)
&=-i m^2\frac{(m-i\alpha)}{(\alpha+\bar\alpha)} \frac{\mu}{8\pi G}
\end{align}
From this last expression, we can compute the extension
\begin{align}
\bar K_{m,-m} &= -i\int_{\partial\Sigma} \Big(i_{\bar\xi_m}
\Delta_{\hat{\bar{\xi}}_{-m}} \ell - i_{\bar\xi_{-m}}
\Delta_{\hat{\bar{\xi}}_m} \ell \Big) \\
&= 
\frac{A}{4\pi G(\alpha+\bar\alpha)} m^3.
\end{align}
As before, the $\bar L_n$ generators are then seen to satisfy
a Virasoro algebra with central charge
\beq\label{eqn:cbar}
\bar c = \frac{3A}{\pi G(\alpha+\bar\alpha)},
\eeq
which is the same value as $c$ given in (\ref{eqn:c}). 
Note that $c, \bar{c}$ given in (\ref{eqn:c}), (\ref{eqn:cbar}) 
are twice the values computed in \cite{Chen:2020nyh, Haco2018}.  
This factor of 2 will have an effect on the entropy computed
in section \ref{sec:canonCardy}.

\subsection{Frame dependence}
Although the null normal is fixed to coincide with the Killing 
horizon generator in the definition of the near-horizon phase
space, we would like to understand how 
the central charges depend on the choice of background
scaling frame.  This is relevant because the choice of frame
was related to the choice of stretched horizon in 
section \ref{sec:stretched}, and since this frame has parallels to a choice of Weyl frame in a CFT, we would like the central charge to be insensitve to this choice. 
Under the rescaling transformation 
(\ref{rescale}), 
the parameter $w_\xi$ characterizing the noncovariance of $l_a$ 
transforms according to 
\beq
w_\xi \rightarrow w_\xi + \lie_\xi f.
\eeq
Using (\ref{bdyanomaly}), this then leads to a change in the anomaly of the boundary term by
\beq
\Delta_{\hat\xi} \ell \rightarrow\Delta_{\hat\xi}\ell -\frac{\eta}{8\pi G} \lie_l \lie_\xi f.
\eeq
For the $\xi_n^a$ generators on $\mathcal{H}^+$, this results in an extra contribution to $K_{m,-m}$ given by the integral over the bifurcation
surface of the following quantity: 
\beq
\frac{\mu}{2\pi G} \frac{m}{(\alpha+\bar\alpha)^2} 
\left[\alpha(m^2+\bar\alpha^2)V \frac{\partial f}{\partial V}+
\frac{\partial}{\partial\phi} 
\left((\alpha\bar\alpha-m^2)f
+(\alpha+\bar\alpha)\frac{\partial f}{\partial V} \right)
\right].
\eeq

The term involving a total $\phi$ derivative integrates to zero,
and hence does not affect the central charge.  The term that can
affect the result is the one proportional to $V\frac{\partial f}{\partial V}$ in the limit $V\rightarrow 0$.  If $f$ is a regular 
function of $V$ at $V=0$, this term drops out and the central
charge is unaffected.  To get a nonzero contribution from 
it, we would need $f\sim \lambda \log V$, corresponding to 
a rescaling of $l^a$ by $V^\lambda$.  This then affects the rate
at which $l^a$ vanishes (or blows up) as the bifurcation surface is approached.  
For example, given the form of $l^a$ in (\ref{eqn:la}),
we see that $\lambda = -1$ rescales $l^a$ to an affine parameterization,
since $V$ is an affine parameter.  

In order to arrive at an unambiguous value of the central charge, we 
must disallow transformations that affect the rate at which $l^a$
vanishes as $V\rightarrow 0$.  This means choosing a normalization
so that it vanishes linearly with respect to an affine parameter
as bifurcation surface is approached, just as 
the horizon-generating Killing vector does.  Note that this still allows for 
rescalings of the generator in a $\phi$ or $\theta^A$-dependent manner,
or, relatedly, making a different choice of 
the affine parameter with respect to 
which $l^a$ vanishes linearly.
However, it rules out using an affinely parameterized generator
when analyzing bifurcate null horizons.  Using the Killing 
parameterization of the null generator is natural
for Killing horizons, but it may be that other choices are preferred
for different setups.  Note that in \cite{Haco2018, Chen:2020nyh},
it seems that a nonstandard choice of this normalization was used,
which happened to set any contribution to the 
central charge from the flux to zero  except the \Hajicek term.  
It would be interesting to explore these other
normalizations in more detail in the future.

\section{Entropy from the Cardy formula}
The relevance of equations (\ref{eqn:c}) and (\ref{eqn:cbar}) 
for the central charges is that they contain information
about the entropy of the horizon.  
To see how this comes about, we need to associate a quantum system with the 
near-horizon degrees of freedom.  It is well known that in a theory
with gauge symmetry such as general relativity, the introduction of a spatial
boundary breaks some of the gauge invariance,  thereby producing additional
degrees of freedom on the boundary 
that would otherwise not have been present \cite{Balachandran_1996,Carlip1995,Donnelly2016a}.  
The edge modes that arise in this fashion are acted on by the quasilocal
charges identified in the previous sections, and thus represent a 
classical system with Virasoro symmetry.  The quantization of this system
should respect the symmetry, and since two dimensional conformal field
theories share this symmetry algebra, we are led to the postulate that the 
quantum system should be a 2D CFT.  In such a theory, the 
asymptotic density of states depends in a universal way on the central
charge according to the Cardy formula \cite{Cardy1986}.  We will find that 
applying this formula in the context of a Killing horizon shows that the 
entropy of the CFT is directly related to the entropy of the horizon.  

\subsection{Canonical Cardy formula} \label{sec:canonCardy}
The Cardy formula comes in two flavors: microcanonical and canonical.  The 
canonical formula applies to a CFT in a thermal state at high temperatures,
and states that the entropy is given by
\beq\label{eqn:Scardy}
S_\text{Cardy} = \frac{\pi^2}{3}(c\, T + \bar c\, \bar T),
\eeq
where $T$ and $\bar T$ are known as the left and right temperatures;  they
are the thermodynamic potentials conjugate to the $L_0$ and $\bar L_0$
charges.  

To apply this formula in the context of a Killing horizon, 
we need to identify the temperatures.  
This can be done in a manner similar to the determination of the  
Hawking temperature in terms of the horizon surface gravity.  
We would expect the density matrix for quantum fields 
just outside of the horizon to be in the Frolov-Thorne
vacuum \cite{Guica2009,Frolov1989, Compere2012}, which is thermal with respect to the horizon-generating Killing 
vector $l^a$.  This means the density matrix should be of the form
\beq
\rho \sim e^{-\frac{2\pi}{\kappa} \omega_l},
\eeq
where $\omega_l = -k_a l^a$ is the frequency of a mode with wavevector $k_a$, 
relative to $l^a$, and the coefficient
$\frac{2\pi}{\kappa}$ is the inverse Hawking temperature.  
Since $l^a$ can be expressed in terms of the left and right Virasoro
vector fields via $\frac{1}{\kappa} l^a = \frac{1}{\alpha}\xi_0^a 
+ \frac{1}{\bar\alpha} \bar\xi_0^a$, the density matrix can equivalently 
be written
\beq
\rho\sim e^{-\frac{2\pi}{\alpha} \omega_0 -\frac{2\pi}{\bar\alpha} \bar\omega_0}
\eeq
where now $\omega_0 = -k_a \xi^a_0$, $\bar\omega_0 = k_a\bar\xi^a_0$
are the frequencies with respect to the Virasoro zero mode generators.  
This then leads us to identify the left and right temperatures
\beq
T = \frac{\alpha}{2\pi}, \qquad \bar T = \frac{\bar\alpha}{2\pi}. 
\eeq

With these temperatures in hand, the Cardy formula (\ref{eqn:Scardy})
applied using the 
computed values (\ref{eqn:c}), (\ref{eqn:cbar}) for $c, \bar c$ yields
\beq\label{eqn:Scanon}
S_\text{Cardy} = 2\left(\frac{A}{4G}\right).
\eeq
Somewhat unexpectedly, we arrive at {\it twice} the entropy of the horizon. 
To interpret this result, recall that the central charges were computed 
using the Barnich-Troessaert bracket of quasilocal charges.  This bracket
was employed because the quasilocal charges are not integrable, since they
are associated with evolution up the horizon, during which symplectic
flux leaks out.  In order to justify such a calculation, one should introduce an auxiliary system that collects the lost symplectic
flux, allowing integrable generators and Poisson brackets to be defined 
on the total system.  Since we postulated that the edge modes on one
side of the horizon are described by a 2D CFT, it is equally natural
to assume that the auxiliary system is another copy of the same CFT, associated
with edge modes on the other side of the horizon. 
This is the picture that would appear when cutting
a global Cauchy surface for the full spacetime across the bifurcation
surface, in which case the left wedge and its edge modes are the only
additional degrees of freedom in the space, and hence must comprise the 
auxiliary system that collects the fluxes from the right wedge.  
If we assume that the 
Barnich-Troessaert bracket computes the central charge of the total system, 
we would arrive at twice the value of the central charge for one of the CFTs.  
This would explain the appearance of the factor of 2 in (\ref{eqn:Scanon}),
since it is counting the entropy associated with edge modes on 
both sides of the horizon.  If we then traced out the auxiliary
system, we would expect the entropy to be exactly half the value
computed above, and hence would arrive at the correct horizon entropy,
\begin{align} S = \frac{A}{4G}. \end{align}  

This conjectural resolution will be expanded upon in 
section \ref{BTdisc}.   
In order to support this interpretation by way 
of contrast, we turn now to a case where the quasilocal charges are 
in fact integrable, so that no fluxes or auxiliary systems are needed.

\subsection{Integrable charges} \label{sec:integrable}
The other possibility that would produce the correct entropy
is if the boundary term $\ell$ were half the value given in
equation (\ref{bdy}).  This would correspond to different boundary
conditions than Dirichlet, since the flux would now contain
an additional contribution proportional to $\delta k$. 
Although this appears unnatural from the perspective of 
gluing subregions discussed in section \ref{sec:quasicharge}, 
if we were only interested in integrable charges so that the subregion
could be treated as a closed system, any boundary condition that 
results in integrability is valid.  In this section, we will show that such
modified boundary conditions are necessary if demanding that the 
HHPS charges be integrable. 

A useful property of the Barnich-Troessaert bracket is 
that if boundary conditions are imposed 
to make the charges integrable, it reduces to the
Dirac bracket of these charges on the submanifold of phase space defined 
by imposing the boundary conditions as constraints.  
The integrable charges therefore 
need not be considered quasilocal, 
but rather are legitimate Hamiltonians generating the 
symmetry on the constrained phase space.  Note, however, that 
the vector fields generating the symmetry must preserve the boundary condition
imposed, i.e.\ they must be tangent to the constraint submanifold, since 
otherwise they do not produce well-defined transformations of the constrained
fields.  

Finding  a boundary condition that ensures vanishing symplectic flux
but is also preserved by the vector fields (\ref{eqn:xin}) and (\ref{eqn:barxin}) is somewhat nontrivial, since the vector fields tend to 
violate any local condition fixing the intrinsic 
or extrinsic quantities on the horizon, 
see equations (\ref{eqn:delxik}), (\ref{eqn:delxil}), (\ref{eqn:delbxik}), and (\ref{eqn:delbxil}).  
However, as discussed in 
\cite{Chen:2020nyh}, one can consider more general conditions 
that are preserved by the symmetry generators, involving
integrals of variations of quantities over portions of the 
horizon.  Assuming such 
a condition is found, the fact that the fluxes then vanish consequently implies that 
the bracket $\{L_n, L_{-n}\}$ can be computed simply from contracting the 
vector fields $\hat\xi_n$, $\hat\xi_{-n}$ into the symplectic form $\Omega$.\footnote{As discussed in section \ref{sec:BTbrack}, 
the central charge is independent of the choice of corner term
$\beta$.}
This computation was already performed in \cite{Chen:2020nyh}, 
and the resulting central charges are 
\begin{align}
c &= \frac{24}{(\alpha+\bar\alpha)^2} \left(\frac{\bar\alpha A}{8\pi G}+
J_H\right) \\
\bar c &= \frac{24}{(\alpha+\bar\alpha)^2} \left(\frac{\alpha A}{8\pi G} 
-J_H\right). 
\end{align}

On the other hand, the general formula (\ref{eqn:Kxizeta}) for the extension
in terms of $\Delta_{\hat\xi} \ell$ still remains valid, albeit
with a possibly different choice of boundary term 
than $\ell=\frac{-k}{8\pi G} \eta$. The simplest generalization 
is to take
\begin{align}\ell = \frac{-a k}{8\pi G} \eta,\end{align} with $a$ some constant.  In order to ensure that the values of $L_0$ and $\bar L_0$
are the same when computed on either the future or past horizon,
we must then choose the boundary term on the past horizon to be 
$\frac{ak}{8\pi G} \eta$.  Doing this produces
the central charges 
\begin{align}
c = \bar c = \frac{3aA}{\pi G(\alpha+\bar\alpha)}.
\end{align}
Equating the above two expressions for $c$ and $\bar c$ yields the 
conditions 
\begin{align} \label{eqn:intcond}
\alpha-\bar\alpha = \frac{16\pi G J_H}{A},\qquad
a = \frac12.
\end{align}
The first condition restricts the parameters $\alpha,\bar\alpha$ defining 
the symmetry generators, and was identified in \cite{Chen:2020nyh} as 
a necessary condition for integrability of the charges.  The second condition $a=\frac12$
shows that the boundary term $\ell$ is half of the value used when 
imposing a Dirichlet flux condition.  It implies that the central charges are 
now half of the value computed in section \ref{sec:centralcharges},
\beq \label{eqn:cint}
c = \bar c = \frac{3A}{2\pi G(\alpha+\bar\alpha)},
\eeq
and consequently the entropy coming from the canonical Cardy formula 
(\ref{eqn:Scardy})
now agrees with the horizon entropy,
\beq \label{eqn:Scanonint}
S_\text{Cardy} = \frac{A}{4G}.
\eeq

\subsection{Microcanonical Cardy formula} \label{sec:microcanonical}
The canonical Cardy formula requires the left and right temperatures
as inputs, which were identified for the horizon using properties
of the Frolov-Thorne vacuum for quantum fields outside of the horizon. 
A more microscopic derivation of the entropy would utilize the microcanonical
Cardy formula, which expresses the entropy in terms of the density 
of states at fixed, large values of $L_0$, $\bar L_0$.  The microcanonical
expression for the entropy is 
\beq\label{eqn:Smicro}
S_{\mu\text{Cardy}} = 2\pi\left(\sqrt{\frac{c L_0}{6}} +\sqrt{\frac{c\bar L_0}
{6}} \right).
\eeq

To apply this formula, we need the values of the charges $L_0$ and $\bar L_0$.
Note that we should  expect the microcanoncial formula to work only in the 
case that the charges are integrable, since only then do 
$L_0$, $\bar L_0$ represent global charges for a closed system.
This is consistent with standard thermodynamics, in which the 
microcanonical ensemble counts the number of states within 
a fixed energy band of a closed system, while the canonical 
ensemble is used for an open system interacting with a bath
at fixed temperature.

According to the discussion in section \ref{sec:integrable}, integrability of 
the charges requires that the boundary term $\ell$ be on future horizon
\beq\label{eqn:ellk}
\ell = -\frac{k \eta}{16\pi G} ,
\eeq
and the past horizon expression is just $\ell = \frac{k\eta}{16\pi G}$, which are half the values they take under Dirichlet flux matching. 
This boundary
term enters explicitly into the expression for the charges via
equation (\ref{eqn:Hxi}), and making the choice (\ref{eqn:ellk})
is important for 
finding the right entropy from the microcanonical Cardy formula.

Including the contribution from the boundary term (\ref{eqn:ellk}), 
we now find that the zero mode charges are 
\begin{align}
L_0 &= 
\frac{\alpha}{\alpha+\bar\alpha}
\left(\frac{\bar \alpha A}{16\pi G} + J_H\right) = 
\frac{\alpha^2}{(\alpha+\bar\alpha)} \frac{A}{16\pi G} \\
\bar L_0 &= \frac{\bar\alpha}{\alpha+\bar\alpha} 
\left(\frac{\alpha A}{16\pi G} - J_H\right) 
=\frac{\bar\alpha^2}{(\alpha+\bar\alpha)} \frac{A}{16\pi G},
\end{align}
where the latter equalities in these equations
employ the  integrability condition
(\ref{eqn:intcond})
determining $\alpha-\bar\alpha$.   Using these values in the microcanonical
Cardy formula (\ref{eqn:Smicro}) with the central charges (\ref{eqn:cint})
gives
\beq
S_{\mu\text{Cardy}} = \frac{A}{4G},
\eeq
in agreement with the canonical result
(\ref{eqn:Scanonint}) and coinciding with the horizon entropy.

\section{Discussion}\label{sec:discussion}

In this work, we revisited the Wald-Zoupas construction of 
quasilocal charges and fluxes for subregions with null boundaries,
with the goal of systematically deriving the central charges that have appeared
in several recent works on symmetries near Killing 
horizons \cite{Haco2018, Haco2019, Aggarwal2020, 
Averin2020, Chen:2020nyh,Perry2020}.  This required generalizing the 
treatment in \cite{CFP} of the Wald-Zoupas procedure for null boundaries by allowing for the most general boundary conditions
consistent with the presence of a null hypersurface. In the process,
we arrived at a general formula (\ref{eqn:Kxizeta})
for the algebra
extension that appears in the quasilocal charge algebra, which
would be applicable in other investigations of near horizon
symmetries. We showed that the central charge arises from fixing $l_a$ as the background structure, 
which we related to a choice of  stretched horizon.
In this context, the central charge arises as an anomaly, 
in a manner quite analogous to the holographic 
Weyl anomaly appearing in AdS/CFT due to noncovariance of the 
gravitational action under changes in the 
radial cutoff. Applying
the Cardy formula to the central charges of a bifurcate, axisymmetric Killing horizon obtained using the 
Dirichlet flux condition yielded twice the entropy of the horizon,
and we argued that the factor of $2$ could be indicative of a 
complementary set of edge modes on the other side of the horizon.  
We now expand upon the possible significance of these results, and end with some future directions.

\subsection{Algebra extension as a scaling anomaly}\label{discussanomaly}
The formula (\ref{eqn:Kxizeta}) for the algebra extension
$K_{\xi,\zeta}$ shows that extensions only arise when the 
boundary term $\ell$ is not covariant with respect to the 
transformations generated by $\xi^a$, $\zeta^a$.  In several other 
treatments of symmetries at null boundaries, the boundary
term was chosen to be covariant, and equation
(\ref{eqn:Kxizeta}) therefore explains the 
vanishing of the central extensions in those cases 
\cite{CFP,Hopfmuller2018, Adami2020a}.  The fact that the extension
is always of the form of a trivial field-dependent 
cocycle \cite{Barnich:2011mi}
 given by 
equation (\ref{eqn:trivcocycle}), means that 
the boundary term can always be chosen to be covariant so as to
eliminate the extension $K_{\xi,\zeta}$.  However, such a choice
is in conflict with the Dirichlet form of the flux, and hence describes
a physically different setup.  Put another way, there is nontrivial
physics in the choice of boundary term, and we should not
view different choices of this term as a type of gauge freedom.  

By imposing the Dirichlet flux condition, we were inevitably led
to fluxes and boundary terms that were not covariant under the 
boundary symmetries.  This noncovariance seems to be 
a feature, rather than a bug, as it gives 
rise to the central charge which ultimately accounts for the horizon
entropy.  The source of noncovariance came from fixing a  
choice of the null normal $l_a$.  This can be viewed as a choice 
of frame, since there is generally
no preferred normalization of $l_a$ when the surface is null. 
The choice of $l_a$ bears resemblance to the choice of 
radial cutoff when describing asymptotic symmetries, or, equivalently,
the choice of conformal factor when dealing with the conformal
compactification. In holographic renormalization, 
the appearance of conformal anomalies in the dual CFT is known
to be related to anomalous transformations of boundary 
terms in the gravitational action with respect to the radial
cutoff \cite{Henningson1998a, Balasubramanian1999a, DeHaro2001, 
Papadimitriou2005, Ciambelli2020}.  
Changing the radial cutoff then affects the induced metric
in the limit that the conformal boundary is approached, and hence
coincides with a choice of Weyl frame in the CFT.

To strengthen the analogy between this notion of conformal
frame and the scaling frame of $l_a$, we showed 
in section \ref{sec:stretched} that a preferred 
normalization of $l_a$ is determined if one specifies a 
sequence of stretched horizons that asymptote to the null surface. 
As has been remarked before, there are multiple ways to 
stretch the horizon \cite{Carlip2011a}, and here we 
see that this ambiguity has a precise analog in terms of the 
scaling frame of $l_a$.  Furthermore, the 
ambiguity in stretching the horizon, or equivalently, choosing the 
scaling frame of $l_a$, is actually responsible
for the appearance of the central charges in the horizon symmetry
algebra.  
The radial vector $\rho^a$ introduced in 
equation (\ref{eqn:rhoa}) generates transformations
that change the stretched horizon foliation pointwise, acting 
like a dilatation about the bifurcation surface.
Intriguingly, we showed in section \ref{sec:noncovariance} 
that the 
$\rho^a$ component of the symmetry generators is
solely responsible 
for producing anomalous transformations of objects on the horizon.
This suggests that $\rho^a$ should be thought of as generating 
changes in the scaling frame of the horizon CFT, just as the 
radial vector in AdS generates Weyl transformations for 
the holographic CFT.  
The central charge in the horizon quasilocal charge algebra 
appears as a classical 
diffeomorphism anomaly coming from $\Delta_{\hat\xi} \ell$,
and experience
with holographic anomalies tells us that it should be interpreted
as a quantum anomaly in a dual quantum description 
 \cite{Henningson1998a, Balasubramanian1999a, DeHaro2001, 
Papadimitriou2005}.  The Virasoro central charge 
indeed has this interpretation in 2D CFTs, where
it appears as an anomaly in the CFT stress tensor 
\cite{difrancesco1997}.  

The interpretation of the central charge as an anomaly
may help explain why computations involving the Cardy formula 
do such a good job of capturing the black hole entropy.  It is somewhat
surprising that a set of Virasoro symmetry generators appear
for Killing horizons of arbitrary dimension, when standard
holographic reasoning would suggest that 
a higher dimensional CFT should appear for higher dimensional black
holes. It is also surprising that seemingly disparate symmetry 
algebras, including $\text{BMS}_3$ \cite{Bagchi2013,
Carlip2018a, Carlip2020}, Virasoro-Ka\v{c}-Moody
\cite{Aggarwal2020}, Heisenberg \cite{Afshar2016}, or just a single copy of Virasoro \cite{Carlip_1999, Carlip_1999a}, all seem to reproduce
the black hole entropy when a Cardy-like formula is available, 
even though each of these symmetries would coincide with physically
different quantum theories.  Some insight into this situation
comes from recalling that the Cardy formula is derived using the 
anomalous tranformation of the stress tensor when performing a change
in conformal frame from the plane to the cylinder 
\cite{Cardy1984,
Cardy1986}.  
The conformal anomaly  determines the vacuum expectation 
value of the stress tensor, which is attributed to a Casimir 
energy associated with putting the theory on a cylinder.  Modular
invariance then relates this vacuum energy to the high temperature 
density of states, from which one arrives at the Cardy formula
for a CFT.  The central charge appears in this 
formula in its capacity as an anomaly coefficient, and it 
may be that this conformal anomaly controls the density of 
states in more general contexts when an exact 2D CFT description 
is not valid.\footnote{For example, a version of the Cardy formula
for higher-dimensional CFTs 
was derived in \cite{Shaghoulian:2015kta}.}  In such a scenario, the extension in the 
quasilocal algebra would continue to characterize the rescaling
anomaly, and one might hope that a suitable generalization
of the Cardy formula would still reproduce the black hole 
entropy.  
Note, however, that modular invariance is a crucial input in the derivation
of the Cardy formula, and hence it should play an important role in 
arriving at the correct entropy.  

\subsection{Barnich-Troessaert bracket and Dirichlet matching}\label{BTdisc}

The Barnich-Troessaert bracket given in (\ref{eqn:BTbracket})
played an important role in defining the algebra satisfied
by the quasilocal charges.  As of yet, however, there is no 
derivation of this bracket from first principles.  The main technical problem is in coming up with an object which replaces
the Poisson bracket when dealing with an open subsystem, which 
can lose symplectic flux through a boundary.  There has 
been some work addressing this problem for general phase 
spaces with boundaries \cite{Lewis1986, 
Soloviev1993, Bering2000, Soloviev2000}, 
but it remains to be seen exactly the connection
between these works and the present context of quasilocal charges
in gravity.  
The heuristic derivation of the bracket in section \ref{sec:BTbrack}
 describes how it might arise by including an auxiliary
system which collects the lost symplectic flux, but it would 
clearly be interesting to carry out such a construction in 
full detail.  

A step toward deriving the Barnich-Troessaert
bracket was taken
by Troessaert in \cite{Troessaert:2015nia}, who interpreted the 
quasilocal symmetry transformations in terms of a family of 
phase spaces parameterized by a set of boundary sources.  These 
boundary sources are simply the 
values taken by the fields appearing in the flux. For the Dirichlet
form of the flux the, intrinsic metric $q_{ij}$ and null generator
$l^i$ constitute the sources.  This interpretation is inspired
by holography, where the holographic dictionary relates 
boundary values of the fields to sources in the dual CFT, and their
conjugate momenta to expectation values of the sourced
operators \cite{Gubser1998a, Witten1998b}.  
In this case, the momenta $\pi^{ij}$ and $\pi_i$ 
from equations (\ref{eqn:piij}) and (\ref{eqn:pii})
should have the interpretation
of the holographic stress tensor for the null boundary,
similar to the Brown-York stress tensor on the timelike 
boundary in standard examples of AdS/CFT
\cite{Balasubramanian1999a}.
Dirichlet conditions also play an important role in holography, since 
other boundary conditions can lead to conformal field theories
with fluctuating sources or metrics, whose interpretation
as a well-defined theory is less clear \cite{Compere2008}.  
Troessaert describes the 
quasilocal symmetries as ``external symplectic symmetries,'' which
are transformations that act on the boundary sources as well as 
the dynamical fields, and demonstrates that the Barnich-Troessaert
bracket arises in a natural way on this enlarged phase space. External symplectic symmetries have also appeared in the 
context of asymptotically flat spaces, where superrotations
have been shown to be of this character \cite{Compere2016}.

The interpretation of the Barnich-Troessaert bracket in terms of an enlarged phase space decomposed into smaller phase spaces of fixed Dirichlet field values is similar to the description of fixed area states in holography \cite{Akers_2019,Dong_2019}. Specifically, in the latter construction, a bulk Cauchy slice is split across the Ryu-Takayanagi (RT) surface \cite{Ryu:2006bv, Ryu:2006ef}, and a general state in the gravitational Hilbert space is decomposed into superselection sectors corresponding to area eigenstates of the RT surface, each of which classically corresponds to a fixed Dirichlet boundary condition (albeit for a codimension-two boundary as opposed to a codimension-one boundary). This description in terms of fixed area states was important for reproducing the correct Renyi spectrum of holographic states. The analogue of the external symplectic transformations are operators that belong to neither the algebra of the entanglement wedge nor its complement. In other words, such transformations would not preserve the center. 
Fixed area states appeared earlier in a slightly different context in \cite{Carlip_1995}, where it was argued that the Bekenstein-Hawking entropy arises from summing over all fixed area configurations of a black hole in Euclidean gravity. Therefore, it might not be all that coincidental that we needed to fix the Dirichlet form in the symplectic potential in order to reproduce the Bekenstein-Hawking entropy from the Cardy formula; investigating the connection between the present work and these other works would be an interesting next step. 

Ultimately, the Barnich-Troessaert bracket should arise 
from a Poisson bracket on a larger phase space, consisting of 
a subregion and its complement.  When gluing together 
the two subregion phase spaces to construct the global phase space, 
each choice for the form of the flux $\beom$ corresponds to a 
specific matching of the boundary variables.  As
discussed in section \ref{sec:quasicharge}, the Dirichlet flux is 
used to kinematically match the metric on the dividing surface,
while the discontinuity in momenta $\pi^{ij}$ and $\pi_i$ are 
dynamically set equal to the boundary stress energy
by the combined variational principle for the subregion
and its complement, yielding a version of the  
junction conditions for general relativity \cite{PhysRevD.43.1129,
poisson2004a,
Engelhardt_2019}.  Matching the intrinsic data is preferred 
over matching the momenta, since jumps in intrinsic data lead
to distributionally ill-defined curvatures, which we expect to be 
excluded from the gravitational path integral.  
In a complete derivation of the Barnich-Troessaert
bracket, we therefore expect the Dirichlet flux condition to 
play an important role.

\subsection{Edge modes and the factor of 2 }
A surprising result of this work is the appearance
of the additional factor of $2$ in the central charges
(\ref{eqn:c}), (\ref{eqn:cbar}) and 
entropy (\ref{eqn:Scanon}) 
 when using the Dirichlet flux condition
to define the quasilocal charges.  This hints at the existence 
of a pair of CFTs describing the degrees of freedom near the 
horizon.  The gluing picture described in section \ref{BTdisc} 
supports this interpretation, since in such a description, one
would naturally construct a pair of quasilocal charge algebras
before combining them into a global phase space. Once this 
procedure is carried out, it may be that the Barnich-Troessaert
bracket computes the algebra associated with the global Virasoro
charges of the two CFTs combined, which would lead to a central
charge that is twice the value associated with the single CFT on 
one side.  The canonical Cardy formula then returns the total
entropy assuming the CFT is in a global thermal state, but if 
we are interested in the entropy associated only with degrees 
of freedom outside of the horizon, we would first have to trace out
the additional interior degrees of freedom.  This would have the 
effect of halving the value of the entropy obtained, which leads 
to the correct entropy formula, $S = \frac{A}{4G}$. 

A contrasting setup was analyzed in sections \ref{sec:integrable}
and \ref{sec:microcanonical},
in which the quasilocal charges were specialized to integrable ones.  This required
a different boundary term that resulted in central charges and 
an entropy that were both half the values obtained using 
the Dirichlet flux, and hence correctly gave the horizon entropy.  
Integrability of the charges allows the subregion to be 
viewed as a closed system, in which case the central
charge we compute would have to be associated with only a single
CFT.  A further consistency check in this case was 
agreement with the microcanonical Cardy formula, which holds since
the system is isolated.  The interpretation of the Dirichlet 
matching condition then seems to be that it necessarily entails
a description in terms of an open system, and the Barnich-Troessaert
bracket computes the total central charge associated with 
both sets of quasilocal charges.  On the other hand,
the boundary term necessary for integrable charges seems to be 
associated with one-sided generators, which, at least for the 
special choice of parameters given in equation (\ref{eqn:intcond}),
do not require a gluing construction.  
Of course, it may be that there is some other justification
for using the alternative boundary term over the Dirichlet one
in a gluing construction, and it would be interesting to 
explore this possibility further.

\begin{figure}[t]
\centering
    \begin{subfigure}[b]{0.4\textwidth}
        \centering
        \includegraphics[height=0.25\textheight]{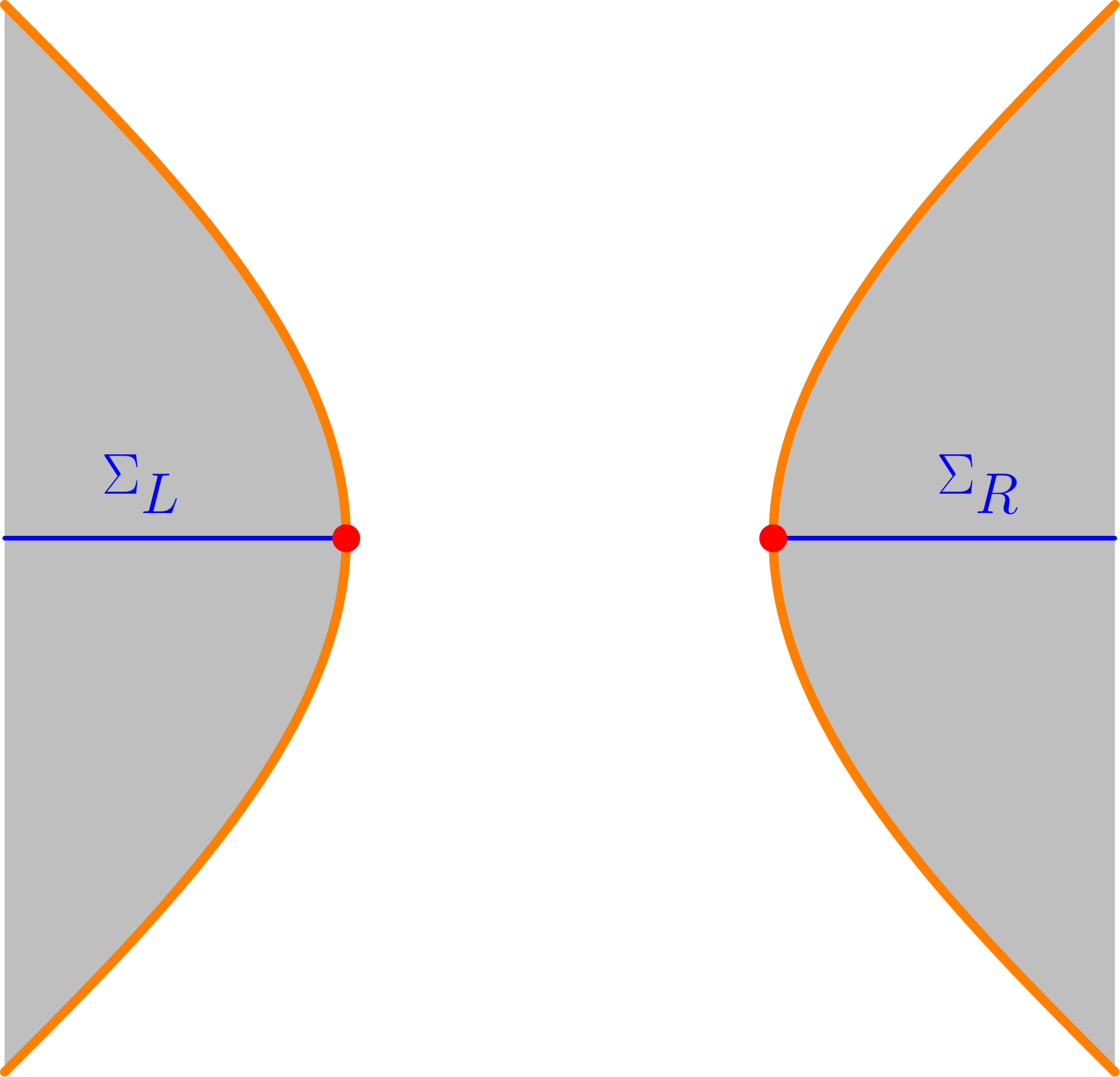}
        \caption{Subregions before gluing}
        \label{fig:before}
    \end{subfigure}
    \hfill
    \begin{subfigure}[b]{0.4\textwidth}
        \centering
        \includegraphics[height=0.25\textheight]{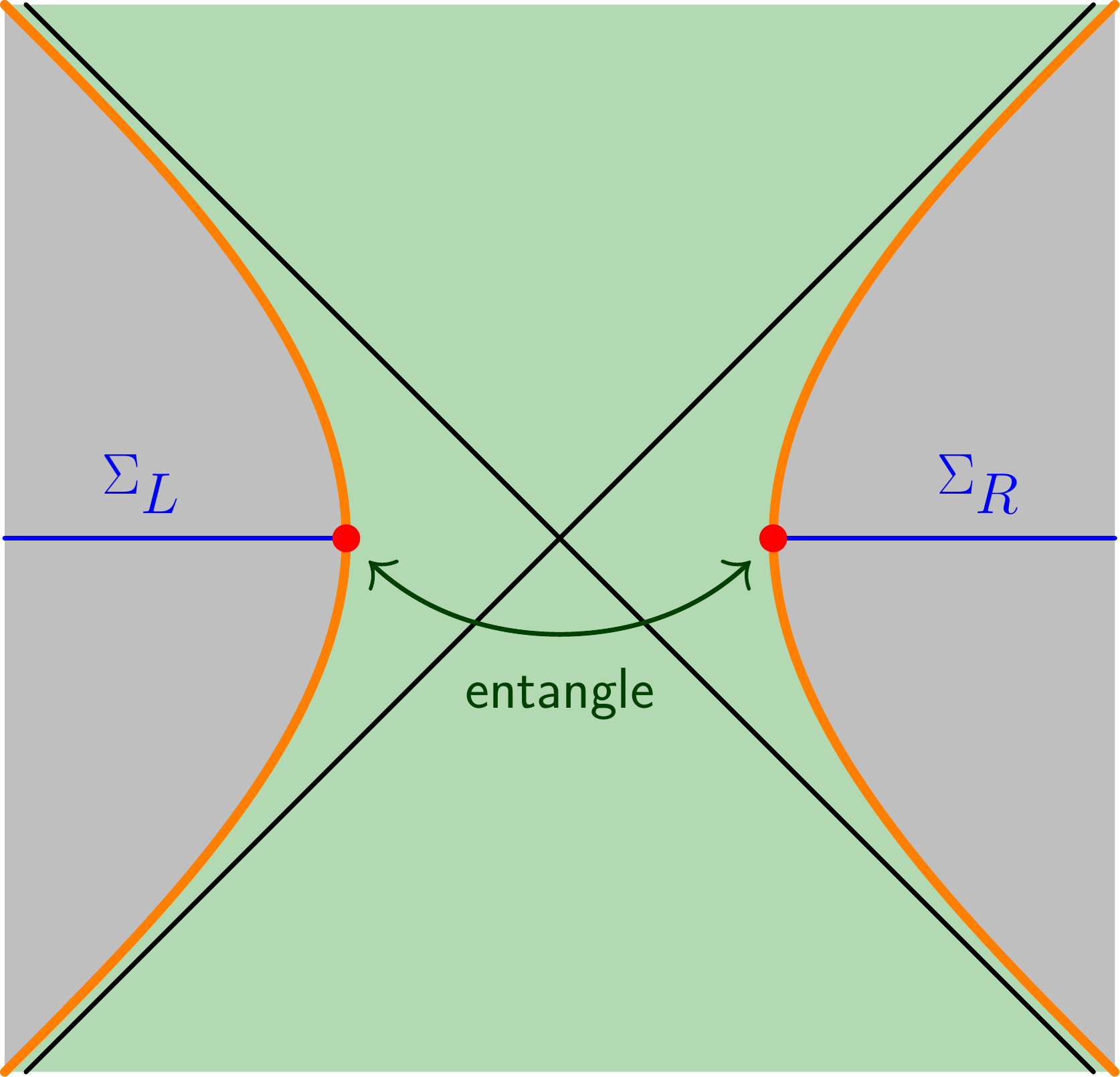}
        \caption{ Connected geometry after gluing }
        \label{fig:after}
    \end{subfigure}
    \caption{Depiction of the gluing procedure.  In (\ref{fig:before}) we show two disconnected subregions, bounded by timelike stretched horizons
    in orange.  
    The boundaries of the respective Cauchy surfaces 
    {{$\Sigma_L$}} and 
    {$\Sigma_R$} are given by the red dots.  In (\ref{fig:after}),
    we imagine gluing the subregions by entangling the edge modes
    on {$\partial \Sigma_L$} 
    with those on {$\partial\Sigma_R$}. 
    This entanglement should build up the geometry of the intervening
    space.  For the nonextremal horizons considered in this paper, 
    the stretched horizons can approach the bifurcate null horizon,
    and the gluing occurs accross the bifurcation surface, with
    the entanglement building up the geometry of the interior.
    }
    \label{fig:gluing}
\end{figure}

This picture in terms of a pair of CFTs arises naturally when
interpreting the horizon entropy as an entanglement entropy.  
In a theory with gauge symmetry such as general relativity, the 
quantum mechanical Hilbert space does not factorize into a tensor
product associated with a subregion and its complement.  However,
one can form an extended Hilbert space \cite{Donnelly:2014gva} 
that does factorize by
introducing additional edge mode degrees of freedom on the boundary
which are acted on by a quasilocal charge algebra closely related
to the ones considering in the present work \cite{Speranza2018a,Donnelly2016a}.
The physical Hilbert space is then identified with a particular
subspace  of the extended Hilbert space, which is constructed 
in a way analogous to the gluing construction described above. This gluing procedure produces entanglement between the edge modes, which ultimately
contributes to the entropy of the state \cite{Donnelly:2014gva}, 
and in some cases 
can be the dominant contribution.  

In the context of this work, since the quasilocal
symmetries contain a Virasoro algebra, we expect each set of 
edge modes to be described in terms of a CFT.  
In order to apply the Cardy formula, this CFT must be modular-invariant, 
which is an additional assumption beyond requiring that the 
edge mode theory  furnish a representation of the Virasoro algebra.  
In fact, if one worked with an irreducible representation of Virasoro,
the density of states would grow like a CFT with central charge $c=1$,
which is clearly insufficient to reproduce the horizon
entropy.\footnote{We thank Alex Maloney for discussions on this point.}
A possible way to view the effect of modular invariance on the edge 
mode description is to think of it as an additional symmetry that 
acts on the edge modes, which then implies additional degeneracy 
of the states when the edge mode theory is quantized. 
This additional dengeneracy coming from modular invariance appears to 
be important for arriving at the correct value of the Bekenstein-Hawking
entropy.

The gluing 
procedure for the edge modes should entangle the pair of CFTs at the 
boundaries into something like a thermofield
double state. 
This creates a picture that is quite familiar
from holography, where entanglement between a pair of CFTs 
builds up a connected black hole 
geometry in the bulk \cite{Maldacena2003b, VanRaamsdonk:2010pw, Maldacena_2013} 
(see figure \ref{fig:gluing}).
The difference when working on the horizon is that when gluing
at the bifurcation surface, the two sets of edge modes are coincident,
as opposed to being spatially separated by the $\text{AdS}$ interior. 
Nevertheless, one might attribute the smooth region to the 
future of the bifurcation surface
as arising from the edge mode entanglement, similar to how smooth
bulk geometries arise from entanglement in holography.  
If one instead worked on the stretched horizons, there would be 
a small spatial region between the gluing surfaces which could 
be thought of as built up from edge mode entanglement.  

In a limit where the horizon approaches extremality with
$\kappa\rightarrow 0$, the stretched horizon picture begins to 
look like standard derivations of holographic dualities
\cite{Maldacena1999b, Aharony2000b}.  The additional ingredient in AdS/CFT
is the appearance of a long $\text{AdS}$ throat, separating 
the stretched horizon from what would have been a bifurcation surface,
were it not infinitely far away.  Associated with this throat is the 
existence of a decoupling limit between modes deep within the 
throat and excitations in the distant asymptotically flat region, 
which allows the CFT dual to the AdS throat to be treated as a closed
system. This decoupling limit is not available for the nondegenerate
horizons considered in this paper, and the CFT associated with the 
quasilocal charges must be thought of as interacting with degrees 
of freedom in the exterior. The need to employ the Wald-Zoupas 
procedure due to the presence of fluxes can be viewed as an indication
of this lack of decoupling.\footnote{Note also that since
we are considering a CFT coupled to an auxiliary system,
it is not immediately clear that the standard Cardy formula still
applies.  It may turn out that this formula is corrected
due to the interactions, and this could yield an alternative
resolution of the factor of $2$ issue.  We thank Tom Hartman for
this suggestion .  }  Although nonstandard in traditional
treatments of AdS/CFT, recent works on black hole evaporation in
holography have employed a similar setup, where the standard 
Dirichlet boundary conditions in AdS are relaxed to allow
fluxes of Hawking radiation to escape into an auxiliary asymptotically
flat region \cite{penington2019entanglement, Almheiri_2019}. Time translation in such a setup should then be viewed as an external symplectic symmetry of the AdS subregion, and the definitions of energy and the boundary symmetry algebra 
would require the Wald-Zoupas procedure and the Barnich-Troessaert bracket. Understanding the quasilocal symmetry algebras of horizons
may therefore provide additional insights into the black hole 
evaporation process and information paradox.

\subsection{Future work}
This work raises a number of questions that motivate further 
investigation. Foremost amongst these is the interpretation of the 
Barnich-Troessaert bracket and its relation to the gluing of subregions.  Deriving the bracket from a gluing construction would make progress towards confirming the conjectured origin of the factor of 2 appearing in the central charge with Dirichlet flux matching.  Beyond that, the gluing construction would demonstrate a way to describe a localized subregion in gravity, from which one could ask additional questions 
about local gravitational observables. On the quantum
side, this gluing procedure gives a way to embed the global gauge-invariant Hilbert space 
of the theory into an extended Hilbert space, and allows notions of 
entanglement entropy for a subregion to be defined. It should also
have a description in terms of the sewing of path integrals
\cite{Carlip_1995, Blommaert2018, Geiller2019}, 
which may also lead to further justifications of the Dirichlet matching
condition.  

Although the main application of this work was an analysis of the Virasoro vector fields for Killing horizons, 
the general formalism we developed
is much more broadly applicable.  In particular,
the expression (\ref{eqn:Kxizeta}) for the central extension in 
terms of the anomalous transformation of the boundary term in the action
applies quite generally, and hence can be utilized for a variety of 
symmetry algebras and types of hypersurfaces.
One interesting application would be to investigate 
the various extended  symmetry algebras that have been proposed for
asymptotically flat space with these methods \cite{Barnich2010e, Barnich2010d,
Campiglia2014b, Strominger2018, Compere2019a}. 
In particular, there may be some connection between the null boundary stress tensor we found in this paper and the celestial 
stress tensor found for 4D asymptotically flat spaces in 
\cite{Kapec2017}, although we expect that suitable counterterms 
to regulate this expression will be needed \cite{Mann2006,
Compere2018}.  It would also be interesting to explore the relation
between these boundary terms and fluxes and the recent work
on effective actions for superrotation modes 
\cite{Nguyen2020}.

More generally, one could look at symmetry algebras associated 
with arbitrary null surfaces \cite{CFP,Adami2020a}, and 
analyze the extensions that appear using the Dirichlet flux condition.  
One intriguing aspect of some of these symmetry algebras is that they
include factors of $\text{Diff}(S^2)$, which is known to have no
nontrivial central extensions.  However, the Barnich-Troessaert bracket
generically produces abelian extensions, which do exist for 
$\text{Diff}(S^2)$.  It would be interesting to see if these 
extensions have any connection to anomalies in a putative quantum
description, and whether one can find a Cardy-like formula related to 
the abelian extensions. 

In \cite{CFP} a BMS-like algebra was found on arbitrary null surfaces, which can be written as a semidirect sum 
$ \text{diff}(S^2) \ltimes\mathfrak{s}$, 
where $\mathfrak{s}$ consists of the generators of the form
$\xi^a = f l^a$. As discussed in section \ref{sec:bcs}, \cite{CFP} employed the boundary condition $\delta k = 0$, which constrains the function $f$ to satisfy $\lie_l(\lie_l + k)f = 0$, so these generators form a pointwise $\mathbb{R}\ltimes \mathbb{R}$ subalgebra corresponding to position-dependent translations and boosts along the null surface, the former of which correspond to supertranslations. We can readily see from our general expression \eqref{eqn:Kxizeta} along with the choice of boundary term \eqref{bdy} on a null surface that the $\delta k = 0$ boundary condition makes the central charge trivially vanish. As explained in \cite{CFP}, if we lift the $\delta k = 0$ condition, then the only modification to the algebra is that now $f$ can be any function on the null surface; such vector fields were considered for example in \cite{Adami2020a}. In particular, if we consider two generators $\xi^a = f l^a$ and $\tilde{\xi}^a = \tilde{f} l^a$, 
the extension $K_{\xi,\tilde\xi}$ computed from (\ref{eqn:Kxizeta})
will be nonzero for an arbitary null surface. A step towards understanding the universality of the Bekenstein-Hawking entropy from the Cardy formula would therefore entail a better understanding of this enlargement of the $\mathbb{R}\ltimes \mathbb{R}$ subalgebra and the resulting abelian extension.

The Wald-Zoupas construction we described in this work required the 
symmetry generators to be tangent to a hypersurface that bounds 
the subregion of interest.  However, diffeomorphisms which move the 
bounding hypersurface should also possess quasilocal charges.  
Treating such transformations would require additional analysis of the 
decomposition of the symplectic potential at the null surface, and a 
characterization of the noncovariances that can arise from such
transformations, but in principle a similar set of techniques should allow
quasilocal charges to be defined for these surface deformations.  
Carrying this out in detail would be a useful next step. 

Another generalization would be to investigate higher curvature 
theories using the Wald-Zoupas procedure. We anticipate this being
more challenging due to the presence of higher time derivatives in the 
action. In particular, we should not expect the Dirichlet flux condition
to be available in general, with the  
exception of Lovelock theories, for which the null boundary 
terms corresponding to Dirichlet conditions are known
\cite{Chakraborty2019}.  
Determining a suitable generalization of that condition would be 
the main obstacle one would need to overcome.
The analysis of \cite{Azeyanagi2009a} on near horizon symmetries of 
extremal black holes in higher curvature theories may
give some insights into this problem.  

Finally, an open question related to the Virasoro symmetry generators considered in \cite{Haco2018} is 
with regards to their geometrical significance.  In the extremal limit,
the generators become symmetries of a warped $\text{AdS}_3$ throat
\cite{Guica2009, Compere2012},
but away from extremality their interpretation is less clear.  
In \cite{Castro2010}, the parameters $\alpha$ and
$\bar\alpha$ were determined by a hidden conformal symmetry of the 
scalar wave equation in the near-horizon region.  Determining 
how this symmetry relates to preservation of the tensor $C_{ab}$ 
defined in (\ref{eqn:Cab})  would lead to further insights on the relation between the near-horizon Virasoro generators and null boundary data.

\section*{Acknowledgements}
We would like to thank Glenn Barnich, 
Steve Carlip, Lin-Qing Chen, William Donnelly,
\'{E}anna Flanagan,
Laurent Freidel, Tom Hartman, Alex Maloney, 
and Dominik Neuenfeld for helpful discussions.  
A.J.S is supported by a postdoctoral fellowship at the 
Perimeter Institute. 
Research at Perimeter Institute is supported in part by the Government of Canada through the Department of Innovation, Science and Industry Canada and by the Province of Ontario through the Ministry of Colleges and Universities. V.C. is supported in part by the Berkeley Center for Theoretical Physics; by the Department of Energy, Office of Science, Office of High Energy Physics under QuantISED Award DE-SC0019380 and under contract DE-AC02-05CH11231; and by the National Science Foundation under grant PHY1820912.

\appendix
\section{Commutation relation for anomaly operator} \label{app:anom}
Here, we give a proof of the relation (\ref{eqn:WZcons}) satisfied
by the anomaly operator $\Delta_{\hat\xi}$.  
By writing out the commutator, we find
\beq \label{eqn:DelDel}
[\Delta_{\hat\xi},\Delta_{\hat\zeta} ] = [L_{\hat\xi},L_{\hat\zeta}]
+[\lie_\xi,\lie_\zeta] -[L_{\hat\xi},\lie_{\zeta}]
-[\lie_\xi,L_{\hat\zeta}]
= L_{[\hat\xi,\hat\zeta]_{\SSS}} + \lie_{[\xi,\zeta]}.
\eeq
Here, $[\hat\xi,\hat\zeta]_{\SSS}$ is the Lie
bracket of vector fields on $\SSS$, and to arrive at the second
equality, we use the fact that $[L_{\hat\xi},\lie_\zeta]$ = 0,
since $\zeta^a$ is field-independent, $\delta\zeta^a=0$. The field
space bracket can be related to the spacetime bracket
simply by contracting with a covariant field $\delta g_{ab}$,
\begin{align}
I_{[\hat\xi,\hat\zeta]_\SSS}\delta g_{ab}
&= L_{\hat\xi}I_{\hat\zeta}\delta g_{ab} 
- I_{\hat\zeta}L_{\hat\xi}\delta g_{ab}
=L_{\hat\xi}\lie_{\zeta} g_{ab} -I_{\hat\zeta}\lie_\xi\delta g_{ab}
=\lie_\zeta \lie_\xi g_{ab}-\lie_\xi\lie_\zeta g_{ab}
= -\lie_{[\xi,\zeta]} g_{ab} \nonumber \\
&
=-I_{\widehat{[\xi,\zeta]}} \delta g_{ab},
\end{align}
and hence we derive
\beq \label{eqn:SLie}
[\hat\xi,\hat\zeta]_\SSS = -\widehat{[\xi,\zeta]}
\eeq
for field-independent generators.  Applying this to (\ref{eqn:DelDel})
yields the desired identity
\beq\label{eqn:WZapp}
[\Delta_{\hat\xi},\Delta_{\hat\zeta}] = -\Delta_{\widehat{[\xi,\zeta]}}.
\eeq
It is also useful to note the commutators with $L_{\hat\zeta}$ and 
$\lie_\zeta$,
\begin{align}
[\Delta_{\hat\xi},L_{\hat\zeta}] &= -L_{\widehat{[\xi,\zeta]}}\\
[\Delta_{\hat\xi},\lie_{\zeta}]&= -\lie_{[\xi,\zeta]}.
\end{align}

\section{Derivation of the bracket identity} \label{app:BTident}
Here, we derive the main identity for the Barnich-Troessaert bracket
and the resulting extension $K_{\xi,\zeta}$.  To be completely
general, we do not assume that $\Delta_{\hat\xi}\beta = 0$.  We first
work with the definition (\ref{eqn:Hxi}) of the quasilocal charges,
so that all of $\Delta_{\hat\xi}\beta$ is contained in the flux.
The Barnich-Troessaert bracket is then
\beq
\{H_\xi, H_\zeta\} = I_{\hat\xi}\delta H_\zeta - \int_{\partial\Sigma}
\Big(i_\xi I_{\hat\zeta}\beom - I_{\hat\zeta} \Delta_{\hat\xi}\beta\Big) 
\equiv \int_{\partial\Sigma} m_{\xi,\zeta},
\eeq
where we have written the final result in terms of a local 
2-form $m_{\xi,\zeta} $ to be integrated.  We can calculate the 
expression for $m_{\xi,\zeta}$ on $\NNN$ as follows:
\begin{align}
m_{\xi,\zeta} 
&=
I_{\hat\xi} \delta Q_\zeta + I_{\hat\xi} i_\zeta \delta \ell 
-I_{\hat\xi} \delta I_{\hat\zeta}\beta -i_\xi I_{\hat\zeta}\p\theta
-i_\xi I_{\hat\zeta}\delta \ell + I_{\hat\zeta} i_\xi d\beta
+I_{\hat\zeta} \Delta_{\hat\xi}\beta 
\nonumber \\
&=
-Q_{[\xi,\zeta]} + i_\xi d Q_\zeta  -i_\xi I_{\hat\zeta}\p\theta 
+i_\zeta\Delta_{\hat\xi}\ell - i_\xi\Delta_{\hat\zeta}\ell 
+i_\zeta\lie_\xi \ell -i_\xi \lie_\zeta\ell
-L_{\hat\xi} I_{\hat\zeta}\beta + I_{\hat\zeta} L_{\hat\xi}\beta
\nonumber \\
&\qquad\qquad
+ d\left(i_\xi Q_\zeta - i_\xi I_{\hat\zeta}\beta\right)
\nonumber \\
&=
-Q_{[\xi,\zeta]} - i_{[\xi,\zeta]}\ell +I_{\widehat{[\xi,\zeta]}}\beta
-i_\xi \Delta_{\hat\zeta}\ell + i_\zeta\Delta_{\hat\xi} \ell
-i_\xi i_\zeta(L+d\ell)
+d\left(i_\xi Q_\zeta +i_\xi i_\zeta \ell 
-i_\xi I_{\hat\zeta}\beta\right)
\label{eqn:mxizeta}
\end{align}
where the first equality used the relation (\ref{eqn:thdecomp}) 
for $\beom$, the second equality expanded the variation of $Q_\zeta$
via
\beq
I_{\hat\xi}\delta Q_\zeta = L_{\hat\xi} Q_\zeta = \lie_\xi Q_\zeta
+ \Delta_{\hat\xi}Q_\zeta = i_\xi d Q_\zeta + d i_\xi Q_\zeta -Q_{[\xi,\zeta]}, 
\eeq
the third equality employed the identities
\beq
i_\zeta\lie_\xi \ell - i_\xi \lie_\zeta \ell =
-i_{[\xi,\zeta]}\ell +i_\xi i_\zeta d\ell + di_\xi i_\zeta \ell
\eeq
and 
\beq
-L_{\hat\xi}I_{\hat\zeta}\beta +I_{\hat\zeta}L_{\hat\xi} \beta
= -I_{[\hat\xi,\hat\zeta]_{\SSS}} \beta = I_{\widehat{[\xi,\zeta]}}\beta
\eeq
where the $\SSS$ Lie bracket $[\hat\xi,\hat\zeta]_{\SSS}$
is related to the spacetime Lie bracket for field-independent generators
by a minus sign according to (\ref{eqn:SLie}).  
By integrating (\ref{eqn:mxizeta}) over $\partial\Sigma$, we 
arrive at the desired identity for the bracket,
\beq
\{H_\xi, H_\zeta\} = -\left[H_{[\xi,\zeta]} 
+\int_{\partial\Sigma}
\Big(i_\xi\Delta_{\hat\zeta}\ell - i_\zeta\Delta_{\hat\xi}\ell\Big) \right]
\eeq
noting that the exact term in (\ref{eqn:mxizeta}) integrates
to zero and $i_\xi i_\zeta(L+d\ell)$ pulls back to zero 
since $\xi^a$ and $\zeta^a$ are tangent to the hypersurface
$\NNN$, so their transverse components to $\partial\Sigma$ must 
be parallel to each other.

Note that if we examine the steps leading to (\ref{eqn:mxizeta}),
we see that the terms involving $\beta$ do not mix with the other 
terms, i.e.\ we have an independent identity involving only $\beta$,
\beq \label{eqn:betaident}
-I_{\hat\xi}\delta I_{\hat\zeta} \beta+ I_{\hat\zeta}i_\xi d\beta
+I_{\hat\zeta}\Delta_{\hat\xi}\beta = I_{\widehat{[\xi,\zeta]}}\beta 
-di_\xi I_{\hat\zeta}\beta.
\eeq
This immediately implies that different choices of $\beta$ in
the decomposition (\ref{eqn:thdecomp}) of $\p\theta$ 
do not affect the algebra
or extension $K_{\xi,\zeta}$.  Stated differently, different choices of 
how to separate off the corner term $\beta$  from the flux $\beom$
correspond to changes in the charges associated with trivial
extensions, $H_\xi \rightarrow H_\xi +\int_{\partial\Sigma}I_{\hat\xi}
(\beta -\beta') $.  This explains why the choice of corner 
term did not enter into the results for the central charges 
reported in \cite{Haco2018, Chen:2020nyh}.  

When utilizing the corner
improvement described in appendix \ref{app:cornerimprove}, the modification
of the charges and bracket amounts to shifting $\{H_\xi,H_\zeta\}$
by the term,
\beq
-\int_{\partial\Sigma}\Big(I_{\hat\xi}\delta\Delta_{\hat\zeta}c
-I_{\hat\zeta}\delta\Delta_{\hat\xi}c\Big)  
\eeq
with $c$ defined by equation (\ref{eqn:betadecomp}).  
Then noting that the integrand can be written
\begin{align}
-L_{\hat\xi} \Delta_{\hat\zeta} c +L_{\hat\zeta}\Delta_{\hat\xi}c 
&= 
-\lie_\xi\Delta_{\hat\zeta} c -\Delta_{\hat\xi}\Delta_{\hat\zeta} c
+\lie_\zeta \Delta_{\hat\xi}c +\Delta_{\hat\zeta}\Delta_{\hat\xi} c
\nonumber \\
&= \Delta_{\widehat{[\xi,\zeta]}} c - i_\xi \Delta_{\hat\zeta}dc
+i_\zeta\Delta_{\hat\xi}dc -d(i_\xi \Delta_{\hat\zeta}c - i_\zeta
\Delta_{\hat\xi}c) 
\end{align}
where we have applied the relation (\ref{eqn:WZapp}).  The first
term is the contribution to improved charge $-H_{[\xi,\zeta]}$,
while the second and third terms correct $K_{\xi,\zeta}$, and the 
last term integrates to zero.  This then leads to the 
expression (\ref{eqn:Kimprove}) for the central charge using the 
corner improvement.

Finally, we verify the cocycle identity (\ref{eqn:Kcocycle}) 
that must be 
satisfied by $K_{\xi,\zeta}$. Using the expression
(\ref{eqn:trivcocycle}) for $K_{\xi,\zeta}$ as a trivial 
field-dependent cocycle, we have 
\beq
I_{\hat\chi} \delta K_{\xi,\zeta}
=
\int_{\partial\Sigma}\Big( 
i_\xi L_{\hat\chi} L_{\hat\zeta} \ell -i_\zeta
L_{\hat\chi}L_{\hat\xi}\ell
-i_{[\xi,\zeta]}I_{\hat\chi}\delta \ell
\Big)
\eeq
Then adding cyclic permutations we get
\beq
I_{\hat\chi}\delta K_{\xi,\zeta} + \text{cyclic}
= 
\int_{\partial\Sigma} \Big(
i_\zeta I_{\widehat{[\chi,\xi]}}\delta\ell 
- i_{[\chi,\xi]}I_{\hat\zeta}\delta \ell
-i_{[\zeta, [\chi,\xi]]}\ell
\Big)
+ \text{cyclic},
\eeq
where we note that the cyclic contributions of the form $i_{[\zeta,[\chi,\xi]]}\ell$ 
actually sum to zero by the Jacobi identity.  They are included to
put the right hand side into the form $K_{[\chi,\xi],\zeta]}+
\text{cyclic}$,
which verifies the cocycle identity (\ref{eqn:Kcocycle}).

\section{Corner improvement} \label{app:cornerimprove}
In deriving the expression (\ref{eqn:Hxi}) for the quasilocal charges,
we assumed that the corner term was covariant, $\Delta_{\hat\xi}\beta = 0$.
Although we will find that for a null surface this condition 
is satisfied, it is still interesting to consider the case where
the corner term is not covariant, as it leads to a useful improvement 
to the expression for the quasilocal charges and the extensions
$K_{\xi,\zeta}$.  Another reason to consider this case is to 
resolve an additional ambiguity that arises in the decomposition
(\ref{eqn:thdecomp}) of $\theta$.  Fixing the form of $\beom$ still 
allows us to make the shifts $\ell\rightarrow \ell + da$, $\beta\rightarrow
\beta + \delta a$.  Under this transformation, the quasilocal charge 
transforms as 
$H_\xi\rightarrow H_\xi - \int_{\partial\Sigma} \Delta_{\hat\xi} a$,
and hence $H_\xi$ is sensitive to this ambiguity if $a$ is not covariant.
Since we are allowing for noncovariance in $\ell$, there is no
reason to assume that $\beta$ and $a$ cannot similarly be constructed 
from noncovariant objects.

To handle the case where $\beta$ is not covariant, we return to 
equation (\ref{eqn:dH+flux}) and find that we need a way to separate
$\Delta_{\hat\xi}\beta$ into a contribution to the charge and a contribution
to the flux.  Similar to how we handled $\theta$, we look for a 
decomposition of $\beta$ at $\partial\Sigma$ of the form
\beq \label{eqn:betadecomp}
\beta = -\delta c + \vep.
\eeq
Note that this decomposition should be made on $\NNN$ without pulling back 
$\beta$ to $\partial\Sigma$.  
In principle we could also include an exact contribution $d\gamma$ in 
the decomposition, but these will always end up integrating 
to zero on $\partial\Sigma$.\footnote{However, this type of contribution
may be relevant when considering surfaces  with 
codimension-3 defects, such as caustics on a null surface, or when
considering singular symmetry generators, such as superrotations
\cite{Compere2016, Adjei2020}. } This decomposition allows us to 
identify $\vep$ with a corner contribution to the  flux, while $c$
is the contribution to the charge.  

The improved quasilocal charge can then be written
\begin{align}
H_\xi &= \int_{\partial\Sigma} \Big(Q_\xi +i_\xi \ell -I_{\hat\xi}\beta
-\Delta_{\hat\xi} c\Big) \\
&= \int_{\partial\Sigma} \Big(Q_\xi -I_{\hat\xi}\vep + i_\xi(\ell +dc)\Big)
\label{eqn:Hxiimproved}
\end{align}
and its variation satisfies an equation similar to (\ref{eqn:Hamflux}),
\beq
\delta H_\xi = -I_{\hat\xi}\Omega +\int_{\partial\Sigma}\Big(i_\xi \beom
-\Delta_{\hat\xi} \vep\Big).
\eeq
The 
continuity equation for the change in the charges between two
cuts of $\NNN$ is 
\beq
H_\xi(S_2)-H_\xi(S_1) = \int_{N^2_1} \Big(I_{\hat\xi}\beom -\Delta_{\hat\xi}
(\ell+dc)\Big),
\eeq
with $F_\xi = \int_{\NNN_1^2} I_{\hat\xi}\beom$ still interpreted 
as the flux, but with an anomalous source now given by $\int_{\NNN_1^2}
\Delta_{\hat\xi}(\ell + dc)$.  Finally, the Barnich-Troessaert bracket is 
defined for these charges as 
\beq
\{ H_\xi, H_\zeta\} = -I_{\hat\xi}I_{\hat\zeta}\Omega
+\int_{\partial\Sigma} \Big( 
I_{\hat\xi}(i_\zeta \beom - \Delta_{\hat\zeta}\vep) -
I_{\hat\zeta}(i_\xi \beom - \Delta_{\hat\xi}\vep) \Big)
\eeq
which again satisfies (\ref{eqn:HxiHzeta}) with the extension given by
\beq\label{eqn:Kimprove}
K_{\xi,\zeta} = \int_{\partial\Sigma}\Big(i_\zeta\Delta_{\hat\xi}(\ell +dc)-
i_\xi \Delta_{\hat\zeta}(\ell + dc) \Big).
\eeq

As before, the ambiguities in the decomposition are 
fixed once we have specified the form of the corner flux term $\vep$.
We expect in this case a Dirichlet condition would fix the 
form of $\vep$, and arguments based on the variational
principle should relate the matching to  
codimension-2 junction conditions, such as those considered
in \cite{Engelhardt_2019}.
Once this is done, the shift, $\beta\rightarrow
\beta + \delta a$ causes $c\rightarrow c -a$, while $\vep$ is invariant.
Hence, the combination $\ell + dc$ is also insensitive to this shift, 
and can be viewed as the improvement to the boundary 
Lagrangian $\ell$ by a contribution from a corner Lagrangian $c$.  
We see that many of the improved formulas are obtained from those of 
previous sections by merely replacing $\ell$ with its invariant form,
$\ell +dc$. 

Note that the formula for the improved quasilocal charges
(\ref{eqn:Hxiimproved}) can be used even in the case that $\beta$ is already
covariant.  This could be useful in cases where one wishes for the corner
flux to depend on the geometry of $\partial\Sigma$, in which case 
it will not be covariant with respect to transformations that move $\partial
\Sigma$, even if $\beta$ originally was.

\section{Checking extension is central} \label{app:checkingcentral}
As discussed in section \ref{sec:BTbrack}, the Barnich-Troessaert
bracket of quasilocal charges generically produces an
abelian extension of the associated algebra of vector fields.  
We found that for the generators $\xi_n^a$ and $\bar\xi_n^a$, 
all of the extensions $K_{m,n}$ vanished in the
Killing horizon background except
for $K_{m,-m}$.  However, the quantities $K_{m,n}$ have nonzero
variations, so in principle their brackets with the $L_n$ generators
could show that the algebra is a nontrivial abelian extension
of the Witt algebra.\footnote{See \cite{Ovsienko1998}
for a classification of 
these abelian extensions.}
Here we will demonstrate that in fact the extension
is central, verifying that the resulting algebra is the Virasoro 
algebra.  

The quantity to compute 
for $\chi^c$, $\xi^c$ and $\zeta^c$ three of the $\xi_n^a$ generators
is (ignoring factors of $8\pi G$)
\begin{align}
I_{\hat\chi}\delta \Big( i_\xi  \Delta_{\hat{\zeta}}\ell\Big)
=
-i_\xi I_{\hat\chi} \delta\big(\eta l^c\big)\nabla_c w_\zeta
\end{align}
since $\delta w_\zeta = 0$, which follows from
$\delta w_\zeta l_a = -\delta \Delta_{\hat\zeta} l_a =  \Delta_{\hat\zeta} \delta l_a = 0$. Then
we have 
\begin{align}
I_{\hat\chi} \delta\big(\eta l^c\big) 
&= 
\lie_\chi\big(\eta l^c\big)
+ \Delta_{\hat\chi} \big(\eta l^c\big) = (\lie_\chi \eta) l^c + 
\eta \lie_\chi l^c
=
(I_{\hat\chi}\delta \eta) l^c - (\Delta_{\hat\chi}\eta) l^c 
+\eta[\chi, l]^c
\nonumber \\
&=
\eta\Big(
-w_\chi  l^c +  [\chi,l]^c 
\Big)\nonumber\\
&=
-\eta\left(w_\chi l^c +in\frac{\kappa}{\alpha}  \chi^c\right)
\end{align}
using that $\Delta_{\hat\chi}(\eta l^c) = 0$ for any vector 
that preserves the horizon, and $I_{\hat\chi}\delta \eta \hateq 0$
for the Virasoro vector fields.  The last line uses that 
$l^c = \frac{\kappa}{\alpha} \xi_0^c + \frac{\kappa}{\bar\alpha}
\bar\xi_0^c$ to compute the bracket with $\chi^c$, 
and has chosen  $\chi^c = \xi_n^c$.  

Now setting $\zeta^a = \chi_m^a$, and using the 
expression (\ref{eqn:wxW+}) for $w_\zeta$, $w_\chi$, 
we have that 
\begin{align}
\left(w_\chi l^c+in\frac\kappa\alpha \chi^c\right)\nabla_c w_\zeta
&=
-inm^2\frac{\kappa}{\alpha} \big(W^+)^{\frac{i(m+n)}{\alpha} }
+inm^2 \frac{\kappa}{\alpha} \big(W^+)^{\frac{i(m+n)}{\alpha} } 
=0
\end{align}
using that $l^c = \kappa V\partial_V^c$ on $\mathcal{H}^+$
in Kruskal coordinates (\ref{eqn:la}), 
and $\chi^c = 
\alpha \big(W^+\big)^{\frac{in}{\alpha}} \left(W^+\partial_+^c
+\frac{in}{2\alpha} y \partial_y^c\right)$ in  conformal coordinates (\ref{eqn:xinconf}).  This shows that 
the integrand in $I_{\hat\chi} \delta K_{\xi,\zeta}$ vanishes.
According to the definition (\ref{eqn:HK})
for the Barnich-Troessaert bracket of $K_{\xi,\zeta}$ with the 
other generators, we see that this implies that $K_{\xi,\zeta}$ 
commutes with all generators, and hence must be central.  
Thus we arrive at the advertised result, that we have 
the Virasoro algebra as our extension, as opposed to some other
abelian extension.  The analysis on the past horizon for the 
$\bar\xi_n^a$ generator is analogous, and similarly confirms that 
the $\bar L_n$ generators form a Virasoro algebra.

\bibliographystyle{JHEPthesis}
\bibliography{nullcps}

\end{document}